\documentclass[pra,twocolumn,superscriptaddress,amsmath,amssymb]{revtex4-2}
\usepackage{amsmath}
\usepackage{amsfonts}
\usepackage{amssymb}
\usepackage{graphicx}
\usepackage{color}
\usepackage{txfonts}
\usepackage{float}
\usepackage[titletoc]{appendix}
\usepackage[colorlinks={true}]{hyperref}
\usepackage{textcase}
\hypersetup{citecolor={blue}, filecolor={blue}, linkcolor={blue}, urlcolor={blue}}

\begin{document}
\title{Analytical two-pulse control of universal  single-qubit gates in rotational ultracold NaCs molecules}
\author{Qi Chen}
\affiliation{Hunan Key Laboratory of Nanophotonics and Devices, Hunan Key Laboratory of Super-Microstructure and Ultrafast Process, School of Physics, Central South University, Changsha 410083, China}
\author{Hao-Xuan Luo}
\affiliation{Hunan Key Laboratory of Nanophotonics and Devices, Hunan Key Laboratory of Super-Microstructure and Ultrafast Process, School of Physics, Central South University, Changsha 410083, China}
\author{Jin-Kang Guo}
\affiliation{Hunan Key Laboratory of Nanophotonics and Devices, Hunan Key Laboratory of Super-Microstructure and Ultrafast Process, School of Physics, Central South University, Changsha 410083, China}
\author{Qian-Qian Hong}
\affiliation{Hunan Key Laboratory of Nanophotonics and Devices, Hunan Key Laboratory of Super-Microstructure and Ultrafast Process, School of Physics, Central South University, Changsha 410083, China}
\author{Li-Bao Fan}
\affiliation{{Hunan Provincial Key Laboratory of Flexible Electronic Materials Genome Engineering, \\School
of Physics and Electronic Science,\\ Changsha University of Science
and Technology, Changsha 410114, China}}
\author{Chuan-Cun Shu}
\email{cc.shu@csu.edu.cn}
\affiliation{Hunan Key Laboratory of Nanophotonics and Devices, Hunan Key Laboratory of Super-Microstructure and Ultrafast Process, School of Physics, Central South University, Changsha 410083, China}
\date{\today}

\begin{abstract}
Complex control protocols and sensitivity to experimental imperfections have limited the practical implementation of quantum gate operations. Here, we present an analytical framework for universal single-qubit gates using rotational states of ultracold NaCs molecules. By encoding qubits in the lowest rotational energy levels, we employ a first-order Magnus expansion to derive closed-form unitary evolution from an optimized two-pulse sequence. This approach establishes precise amplitude and phase conditions for arbitrary single-qubit rotations, achieving gate fidelities above 0.9999 in numerical simulations. We further demonstrate that complex multi-gate sequences, including phase-locked operations, can be executed with minimal population leakage into auxiliary states. The time-dependent molecular orientation is shown to faithfully encode both the gate truth table and coherence dynamics, enabling practical gate tomography via weak-field polarization detection. Our analytical method is also applicable to other molecules and physical platforms, offering a potential path to high-fidelity, scalable molecular quantum processors.
\end{abstract}
\maketitle
\section{Introduction}
Quantum computing has the potential to overcome the fundamental limits of classical information processing \cite{steane1998quantum}. The quantum bit (qubit) enables exponentially greater computational capacity than classical bits by leveraging superposition and entanglement \cite{preskill1998reliable,ladd2010quantum}. Universal single-qubit gates, which perform arbitrary rotations on the Bloch sphere, are essential for practical quantum algorithms and fault-tolerant architectures  \cite{campbell2017roads}. Achieving high-fidelity, noise-resilient single-qubit gates remains a key challenge for scalable quantum computing  \cite{zhou2020limits}.\\ \indent 
Several physical platforms have been investigated for quantum information processing, including superconducting circuits, semiconductor quantum dots, trapped ions, and neutral atoms. Superconducting circuits are compatible with microfabrication \cite{wendin2017quantum,gu2017microwave,krantz2019quantum} but suffer from finite coherence times  \cite{siddiqi2021engineering,gyenis2021experimental,ganjam2024surpassing}. Semiconductor quantum dots offer promising integration prospects  \cite{maurand2016cmos,veldhorst2017silicon,gonzalez2021scaling} yet remain vulnerable to charge noise \cite{yoneda2018quantum,huang2018spin,shehata2023modeling}. Trapped ions provide excellent coherence  \cite{harty2014high,wang2017single,wang2021single} but impose demanding hardware requirements  \cite{monroe2013scaling,friis2018observation,kwon2024multi}. Neutral atoms support large-scale arrays  \cite{wu2021systematic,manetsch2025tweezer,wu2025quantum} and parallel operations  \cite{levine2019parallel,evered2023high,m2023parallel} but require sophisticated experimental control \cite{graham2019rydberg,henriet2020quantum,bluvstein2024logical}. Polar molecules present a compelling alternative, leveraging their rich internal structure and permanent electric dipole moments to enable strong, long-range interactions essential for multi-qubit logic gates  \cite{demille2002quantum,bao2023dipolar,langen2024quantum,picard2025entanglement,ruttley2025long}.\\ \indent 
Despite these advantages, developing efficient and precise control protocols for universal single-qubit gates in polar molecules remains challenging. Conventional approaches rely on multiple microwave pulses to achieve arbitrary rotations \cite{langen2024quantum}. While effective, these methods often incur higher error rates, prolonged gate times, and heightened sensitivity to experimental imperfections—including amplitude fluctuations and dephasing  \cite{wang2012effect,souza2011robust,ezzell2023dynamical}. These limitations can constrain the gate fidelities required for scalable molecular quantum processors  \cite{cornish2024quantum}.\\ \indent 
In this work, we present an analytical two-pulse control protocol for universal single-qubit gates in ultracold NaCs molecules. Qubits are encoded in the lowest rotational energy levels, and we use a first-order Magnus expansion to derive the unitary evolution from an optimized two-pulse sequence. We establish universality by specifying amplitude and phase conditions that enable arbitrary single-qubit rotations. To assess feasibility, we conduct numerical simulations for ultracold NaCs molecules. Our results show gate fidelities above 0.9999 and demonstrate that complex multi-gate sequences, including phase-locked operations, can be performed with minimal population leakage into auxiliary rotational states. We also analyze the time-dependent evolution of molecular orientation and show that its dynamics provide a direct observable for qubit readout.
\\ \indent 
The remainder of this paper is organized as follows. Section \ref{Sec:Model} describes the theoretical framework. Section \ref{Sec:Results} presents numerical simulations and discusses the control schemes. Section \ref{Sec:Conclusion} summarizes the principal findings.
\section{Theoretical method}\label{Sec:Model}
\subsection{Description of molecular qubit gates}
\begin{figure}[htbp]
    \centering        \includegraphics[width=0.5\textwidth]{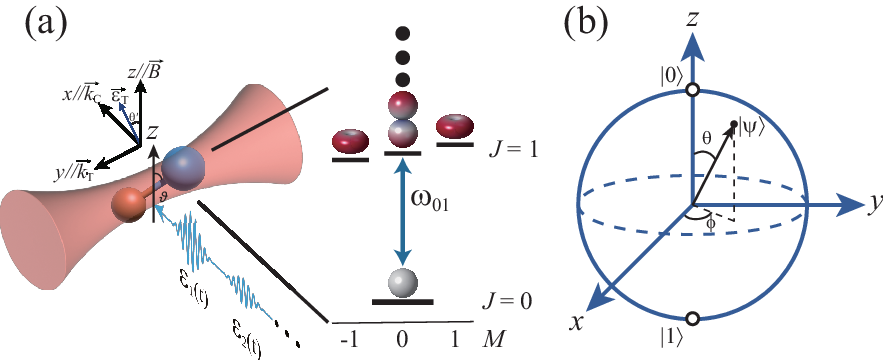}
    \caption{(a) The schematic describing an ultracold linear polar molecule trapped in the optical tweezer being resonantly controlled by the linearly polarized pulse sequence. A bias magnetic field $\vec{B}$ defines the laboratory $z$-axis. Control pulses are $z$-polarized and propagate along the $x$-axis ($\vec{k}_C \parallel \hat{x}$), while the optical tweezer ($\vec{k}_T \parallel \hat{y}$) has its polarization oriented at an angle $\theta'$ relative to $\vec{B}$. $\vartheta$ denotes the orientation angle of the molecular axis relative to the $z$-axis. The qubit is encoded in the rotational subspace $\{|0,0\rangle, |1,0\rangle\}$. (b) The Bloch representation of a single qubit with arbitrary state $ \left|\psi\right\rangle$ and corresponding polar angle $\theta$ and azimuth angle $\phi$.
     }
    \label{figure1}
\end{figure}

Figure~\ref{figure1}(a) shows a pulse sequence applied to ultracold NaCs molecules in the absolute ground state ($X^1\Sigma^+, \nu=0, N=0$), which have a permanent electric dipole moment $\mu_0 = 4.6$~D \cite{aymar2005calculation} and a rotational constant $B = 0.0631$~cm$^{-1}$ \cite{dagdigian1972molecular}. Accurate modeling typically requires including hyperfine structure from the nuclear spins ($I_{\mathrm{Na}} = 3/2$, $I_{\mathrm{Cs}} = 7/2$). However, this can be neglected if either (i) the spectral bandwidth of the driving pulse $\Delta\omega \sim 1/\sigma_t$, or (ii) the peak Rabi frequency $\Omega_{\mathrm{Rabi}} = \mu_0 E_0/\hbar$, exceeds the hyperfine splitting in the $N=1$ manifold ($\delta_{\mathrm{hfs}} \sim 10$--$50$~MHz). The first case applies to broadband, non-selective excitation, while the second corresponds to the strong-drive (Paschen–Back) limit, where the pulse dresses the rotational states and decouples the nuclear spins from molecular rotation. In both regimes, the hyperfine structure is spectrally unresolved, and omitting it introduces negligible error when describing single-photon rotational transitions within the $X^1\Sigma^+$ state. For linear polarization along the quantization axis, the electric-dipole selection rules are $\Delta N = \pm 1$ and $\Delta M_N = 0$. Since $S=0$ in the $^1\Sigma^+$ state, the total angular momentum excluding nuclear spins is $\mathbf{J} = \mathbf{N}$. The first-order Stark shifts induced by the optical trapping light can be eliminated by setting its polarization at the magic angle $\theta' \approx 54.7^\circ$, see  details  in Appendix~\ref{appendix}.
\\ \indent 
\\ \indent 
The corresponding rotational dynamics of the molecule are governed by the Hamiltonian (setting $\hbar = 1$):
\begin{equation} \label{Eq1}
\hat{H}(t) = B\hat{J}^2 - \mu_0 \mathcal{E}(t) \cos \vartheta,
\end{equation}
where the first term represents the field-free rotational Hamiltonian of the diatomic molecule, characterized by the rotational constant $B$ and the angular momentum operator $\hat{J}^2$, with eigenstates $|J M\rangle$ and energies $E_J = B J(J+1)$. The second term describes the interaction between the permanent dipole moment $\mu_0$ and the time-dependent electric field $\mathcal{E}(t)$, where $\vartheta$ denotes the angle between the molecular axis and the pulse polarization direction.\\ \indent 
The unitary evolution operator of the molecules in the interaction picture is given by 
\begin{equation}\label{UTT}
\hat{U}(t, t_0) = \hat{\mathbb{I}}+ i \int_{t_0}^t \mathrm{d}t' \left( e^{i \hat{H}_0 t'} \hat{\mu} \mathcal{E}(t') e^{-i \hat{H}_0 t'} \right) \hat{U}(t', t_0),
\end{equation}
with the initial condition $\hat{U}(t_0, t_0) = \hat{\mathbb{I}}$, where $\hat{\mu}=\mu_0\cos\vartheta$ denotes the molecular dipole operator along the field direction. By considering an $J_{\rm{max}}$-dimensional Hilbert space of rotational states, the field-free Hamiltonian $\hat{H}_0$ in Eq.~(\ref{UTT}) can be expressed as 
\begin{equation}
\hat{H}_0 = \sum_{J=0}^{J_{\rm{max}}}\sum_{M=-J}^J E_J|JM\rangle  \langle JM|,
\end{equation}
and the dipole moment matrix element between the rotational states $|JM\rangle$ and $| (J+1)M \rangle$ is given by
\begin{equation}
\begin{aligned}
\mu_{J,J+1}&= \mu_0 \langle (J+1)M | \cos \vartheta | JM \rangle \\
&= \mu_0 \sqrt{\frac{(J+1)^2-M^2}{(2J+3)(2J+1)}}.
\end{aligned}
\end{equation}\\ \indent 
As shown in Fig.~\ref{figure1}(a), the computational qubit states $|0\rangle$ and $|1\rangle$ are encoded in the rotational states $|0,0\rangle$ and $|1,0\rangle$, respectively, with higher rotational states ($J>1$) serving as leakage channels. By performing the Z-Y decomposition, the unitary operator for a universal single-qubit gate can be expressed as \cite{nielsen2010quantum}
\begin{equation} \label{UG}
\begin{aligned}
 U_{\rm{gate}} &= e^{i\alpha} R_z(\beta) R_y(\gamma)R_z(\delta)\\ &=e^{i\alpha}
\begin{bmatrix}
\cos\left(\frac{\gamma}{2}\right) e^{-i\frac{\beta+\delta}{2}} & -\sin\left(\frac{\gamma}{2}\right) e^{-i\frac{\beta-\delta}{2}} \\
\sin\left(\frac{\gamma}{2}\right) e^{i\frac{\beta-\delta}{2}} & \cos\left(\frac{\gamma}{2}\right) e^{i\frac{\beta+\delta}{2}}
\end{bmatrix},
\end{aligned}
\end{equation}
where $\alpha$, $\beta$, $\gamma$ and $\delta$ are real parameters, and the rotation operators are defined as follows:
\begin{equation}
\begin{aligned}
R_z(\beta)&=
\begin{bmatrix}
e^{-i\beta/2} & 0 \\
0 & e^{i\beta/2}
\end{bmatrix},\\
R_y(\gamma)&=
\begin{bmatrix}
\cos\left(\frac{\gamma}{2}\right) & -\sin\left(\frac{\gamma}{2}\right) \\
\sin\left(\frac{\gamma}{2}\right) & \cos\left(\frac{\gamma}{2}\right)
\end{bmatrix}.
\end{aligned}
\end{equation}
This decomposition establishes a theoretical minimum of three fundamental rotations for the efficient universal implementation of single-qubit gates. In practice, however, direct $R_y$ or $R_z$ operations may require multiple physical controls, thereby increasing the total number of operations. This work aims to design a pulse sequence that implements the three necessary rotations with minimal physical steps while maintaining high fidelity.\\ \indent 
\subsection{Two-pulse control method}
Without employing the rotating wave approximation, the effective Hamiltonian in the interaction picture is given by
\begin{equation}\label{Eq2}
\hat{H}_I(t) = - \begin{bmatrix}
	0 & \mu _{01}\mathcal{E}\left( t \right) e^{-i\omega _{01}t} \\
	\mu _{10}\mathcal{E}\left( t \right) e^{i\omega _{01}t} & 0
\end{bmatrix}.
\end{equation}
To facilitate analytical pulse design, we utilize the first-order Magnus expansion:
\begin{equation}
 \hat{U}_1^{(1)} (t, t_0)= \exp\left\{-i \int_{t_0}^t \mathrm{d}t' \hat{H}_I(t')\right\}.  
\end{equation}
By applying a single pulse $\mathcal{E}(t)=\mathcal{E}_1(t)$, the unitary operator at the target time $t_f$ can be written as 
\begin{equation}\label{U1}
\hat{U}_1^{(1)}(t_f, t_0)=\begin{bmatrix}
 \cos\theta_1 & i\sin  \theta_1 e^{-i\phi_1} \\
 i\sin  \theta_1 e^{i\phi_1} &  \cos  \theta_1 
\end{bmatrix},
\end{equation} 
where $\theta_1= |\Theta_1|$ and $\phi_1= \arg[\Theta_1]$ are the modulus and phase of the complex pulse area, respectively, defined as
\begin{equation}\label{Eq4}
    \Theta_1=\mu_{01}\int_{t_0}^{t_f}\mathrm{d}t'\,\mathcal{E}_1(t')e^{i\omega_{01}t'}.
\end{equation}
Eq.~(\ref{U1}) demonstrates how a single control field $\mathcal{E}_1(t)$ manipulates the qubit state on the Bloch sphere. Specifically, $\theta_1$ determines the rotation angle, and $\phi_1$ specifies the rotation axis in the $xy$-plane, as shown in Fig.~\ref{figure1}(b). Except for the Pauli-X and Y gates, comparing Eq.~(\ref{U1}) and Eq.~(\ref{UG}) shows that a single-pulse operation without spectral phase shaping cannot produce the diagonal phase terms needed for universal single-qubit gates \cite{lee2017single}.\\ \indent 
To overcome this limitation, we propose a two-pulse control scheme that sequentially applies two distinct pulses $\mathcal{E}_1(t)$ and $\mathcal{E}_2(t)$. The combined effect at the final time $t_f$ can be described by the unitary operator
\begin{equation} \label{UT}
     \hat{U}_2^{(1)}=\begin{bmatrix}
 \cos  \theta_{2} &  i\sin  \theta_{2} e^{-i\phi_{2}} \\
 i\sin  \theta_{2} e^{i\phi_{2}}&  \cos  \theta_{2}
\end{bmatrix}\begin{bmatrix}
 \cos  \theta_{1} &  i\sin  \theta_{1} e^{-i\phi_{1}} \\
 i\sin  \theta_{1} e^{i\phi_{1}}&  \cos  \theta_{1}
\end{bmatrix},
\end{equation}
where the rotation angles $\theta_1$ ($\theta_2$) and $\phi_1$ ($\phi_2$) correspond to the modulus and phase of the complex pulse area for the first (second) pulse. To simplify experimental calibration and provide a clear geometric interpretation of universal quantum gate synthesis on the Bloch sphere, we set $\theta_1 = \pi/2$ as a reference pulse. This reduces Eq.~(\ref{UT}) to
\begin{equation} \label{UT1}
     \hat{U}_2^{(1)}=\begin{bmatrix}
 \cos\left(\theta_2+\frac{\pi}{2}\right)e^{-i\left( \phi _{2}-\phi _{1} \right)}&  -\sin\left(\theta_2+\frac{\pi}{2}\right) e^{-i(\phi _{1}+\frac{\pi}{2})} \\
 \sin\left(\theta_2+\frac{\pi}{2}\right)e^{i(\phi _{1}+\frac{\pi}{2})}&  \cos\left(\theta_2+\frac{\pi}{2}\right)e^{i\left( \phi _{2}-\phi _{1} \right)}
\end{bmatrix}.
\end{equation}
This simplification enhances robustness by suppressing common phase noise through relative phase dependence. Comparing Eqs.~(\ref{UT1}) and (\ref{UG}) reveals that, by properly tuning the parameters $\theta_2$, $\phi_1$, and $\phi_2$, the two-pulse protocol allows for the universal implementation of arbitrary single-qubit unitary operations, up to a global phase.\\ \indent 
Table~\ref{tab:gates} summarizes the parameters required to implement key single-qubit gates within the present two-pulse protocol, including the Pauli-Z, Hadamard (H), S, and T gates. Each gate is characterized by its matrix representation and the angular parameters $\alpha$, $\theta_1$, $\phi_1$, $\theta_2$, and $\phi_2$, which define the necessary physical operations. Differences in these parameters reflect the distinct functions of each gate. For instance, the Pauli-Z gate is obtained by introducing a relative phase between $|0\rangle$ and $|1\rangle$, while the Hadamard gate requires a different set of pulse parameters to mix the two computational states. The $S$ and $T$ phase gates introduce relative phase shifts of $\pi/2$ and $\pi/4$.\\ \indent 
\begin{table}[t]
\centering
\caption{Parameters for fundamental single-qubit gate implementation.}
\label{tab:gates}
\setlength{\tabcolsep}{2.5pt}
\begin{tabular*}{\linewidth}{@{\extracolsep{\fill}}lcccccc@{}}
\hline\hline
\\[-4pt]
\textbf{Gate} & $\hat{U}_{\rm{gate}}$ & $\alpha$ & $\theta_1$ & $\phi_1$ & $\theta_2$ & $\phi_2$\\
\hline
Z & $\begin{pmatrix}1 & 0\\ 0 & -1\end{pmatrix}$ & $-\pi/2$ & $\pi/2$ & $-\pi/2$ & $\pi/2$ & 0\\[4pt]
H & $\begin{pmatrix} \sqrt{2}/2 & \sqrt{2}/2\\  \sqrt{2}/2 & - \sqrt{2}/2\end{pmatrix}$ & $\pi/2$ & $\pi/2$ & $-\pi$ & $\pi/4$ & $\pi/2$\\[4pt]
S & $\begin{pmatrix}1 & 0\\ 0 & i\end{pmatrix}$ & $\pi/4$ & $\pi/2$ & $-5\pi/4$ & $\pi/2$ & 0\\[4pt]
T & $\begin{pmatrix}1 & 0\\ 0 & e^{i\pi/4}\end{pmatrix}$ & $\pi/8$ & $\pi/2$ & $-9\pi/8$ & $\pi/2$ & 0\\[2pt]
\hline\hline
\end{tabular*}
\end{table}
\subsection{Pulse design}
Building on our previous works controlling molecular rotations to engineer quantum superpositions (wavefunctions) within the pulse-area theorem, we now extend this approach to quantum precise control of molecular quantum gates (unitary operators) \cite{hong2021generation,fan2023quantum,hong2025precise,fan2025maximizing,hong2026precise,yang2026multilevel}. We focus on designing and optimizing the control pulse $\mathcal{E}(t)$, which consists of two sequential sub-pulses, $\mathcal{E}_1(t)$ and $\mathcal{E}_2(t)$, separated by a controllable time delay $\tau$. In the frequency domain, the Fourier transform of the total field is given by:
\begin{equation}
\int_{t_0}^{t_f}\mathrm{d}t'\,\mathcal{E}(t') e^{-i\omega t'}=A_1(\omega)e^{i\varphi_1(\omega)}+A_2(\omega)e^{i\varphi_2(\omega)}e^{-i\omega\tau},
\end{equation}
where $A_i(\omega)$ and $\varphi_i(\omega)$ denote the spectral amplitude and spectral phase of $\mathcal{E}_i(t)$ for $i=1,2$, respectively. By evaluating this expression at the transition frequency $\omega=\omega_{01}$ and comparing it with the complex pulse areas defined in Eq.~(\ref{Eq4}), we obtain the following relations:
\begin{equation}\label{ME}
 \begin{aligned}
    \theta_{1} e^{-i \phi_{1}} &= \mu_{01} A_{1}(\omega_{01}) e^{i\varphi_{1}(\omega_{01})},\\
    \theta_{2} e^{-i \phi_{2}} &= \mu_{01} A_{2}(\omega_{01}) e^{i\varphi_{2}(\omega_{01})}e^{-i\omega_{01}\tau}.
 \end{aligned}
\end{equation}
These equations establish an explicit mapping between the spectral control parameters—amplitudes $A_1, A_2$, spectral phases $\varphi_1, \varphi_2$, and delay $\tau$—and the gate parameters $\theta_1, \theta_2, \phi_1, \phi_2$ required for the qubit operation in Eq.~(\ref{UT1}).\\ \indent 
Based on the mapping in Eq.~(\ref{ME}), we can synthesize the specific control pulses. As an illustrative example, we consider pulses with Gaussian spectral distributions. The phase-locked spectral field is then constructed as:
\begin{equation}\label{eq17}
E(\omega) = \sum_{i=1}^{2} a_{i} e^{-\frac{(\omega-\omega_{0})^{2}}{2\Delta\omega^{2}}} e^{i\varphi_{i}(\omega)} e^{-i(\omega-\omega_0)\tau_{i}},
\end{equation}
where the amplitudes are set as $a_{i}=\theta_i/\mu_{01}$, and the central frequency is tuned to resonance, $\omega_{0}=\omega_{01}$. To implement the target single-qubit gate, we choose the spectral phases and delays for the two pulses as $\varphi_{1}(\omega)= -\phi_{1}$ (with $\tau_{1}=0$) and $\varphi_{2}(\omega)= \omega_{01}\tau - \phi_{2}$ (with $\tau_{2}=\tau$), respectively. \\ \indent 
Performing the inverse Fourier transform on $E(\omega)$ yields the analytical form of the time-dependent control pulses:
\begin{equation}\label{pulse}
    \mathcal{E}\left(t\right)=\mathcal{E}_1 e^{-\frac{t^2}{2\sigma_t ^2}}\cos \left( \omega _{01}t-\phi_{1} \right)+\mathcal{E}_2 e^{-\frac{\left( t-\tau\right) ^2}{2\sigma_t ^2}}\cos \left( \omega _{01}t-\phi_{2} \right),
\end{equation}
where the electric field amplitudes are given by $\mathcal{E}_1 = \sqrt{2/\pi}\theta_1/(\mu_{01}\tau)$ and $\mathcal{E}_2 = \sqrt{2/\pi}\theta_2/(\mu_{01}\tau)$. The pulse duration (standard deviation) is related to the spectral bandwidth by $\sigma_t = 1/\Delta\omega$. The application of these analytically derived pulses, as specified in Eq.~(\ref{pulse}), enables precise implementation of arbitrary single-qubit operations and ensures that the physical system realizes the desired gate with high fidelity.
\\ \indent 
\begin{figure}[htbp]
    \centering        \includegraphics[width=0.45\textwidth]{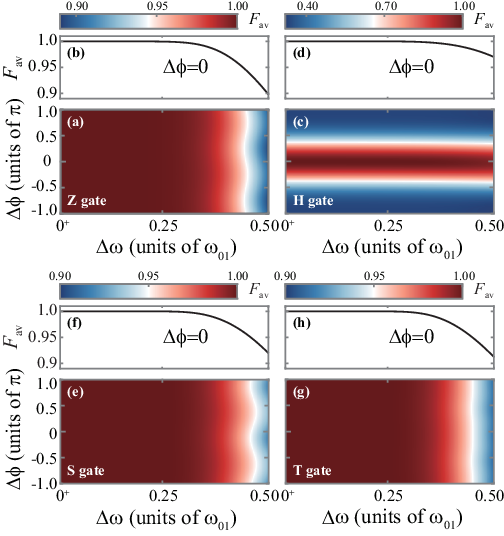}
    \caption{ Average gate fidelity $F_\text{av}$ versus pulse bandwidth $\Delta \omega$ and phase error $\Delta\phi$. Simulations are performed using parameters from Table~\ref{tab:gates}. Lower panels (a, c, e, g) show heatmaps of Pauli-Z, Hadamard, S, and T gates, respectively. The upper panels (b, d, f, h) present the bandwidth-dependent fidelity extracted at zero phase error.}
    \label{fig2}
\end{figure}
\subsection{Average gate fidelity}
To quantify how effectively the implemented unitary operation $\hat{U}$ approximates the target single-qubit gate $\hat{U}_{\text{gate}}$, we employ the average gate fidelity, defined as \cite{pedersen2007fidelity}
\begin{equation}\label{TotalFidelity}
F_\text{av} = \frac{1}{d(d+1)} \left[ \mathrm{Tr}(\hat{M}_{\text{rel}} \hat{M}_{\text{rel}}^\dagger) + | \mathrm{Tr}(\hat{M}_{\text{rel}}) |^2 \right],
\end{equation}
where $d=2$ is the dimension of the single-qubit computational subspace. The matrix $\hat{M}_{\text{rel}}$ characterizes the overlap between the ideal and implemented operations restricted to the computational subspace:
\begin{equation}
\hat{M}_{\text{rel}} = \hat{P} \hat{U}_{\text{gate}}^\dagger\hat{U} \hat{P},
\end{equation}
where $\hat{P}$ is the projection operator onto the qubit subspace. Since the computational basis states $\{|0\rangle, |1\rangle\}$ effectively form a closed system within the interaction space, we adopt $\hat{P} = \hat{\mathbb{I}}_{2 \times 2}$. Consequently, $\hat{M}_{\text{rel}}$ simplifies to $\hat{U}_{\text{gate}}^\dagger \hat{U}_{\text{eff}}$, where $\hat{U}_{\text{eff}}$ represents the top-left $2 \times 2$ block of the full evolution operator $\hat{U}$. This formulation of $F_\text{av}$ represents the probability that the implemented gate $\hat{U}$—which may act on a higher-dimensional Hilbert space ($N>2$)—matches the ideal gate $\hat{U}_{\text{gate}}$, averaged over all pure qubit input states. An optimal fidelity of $F_\text{av} = 1$ indicates perfect gate implementation.
\section{Results and Discussion}\label{Sec:Results}
 To facilitate separate analysis of basic gate-level operations and their integration into larger circuits, the following results and discussion are divided into two subsections: simulations with quantum gates and simulations with quantum circuits. 
\subsection{Simulations for quantum gates}
\begin{figure}[htbp]
    \centering        \includegraphics[width=0.45\textwidth]{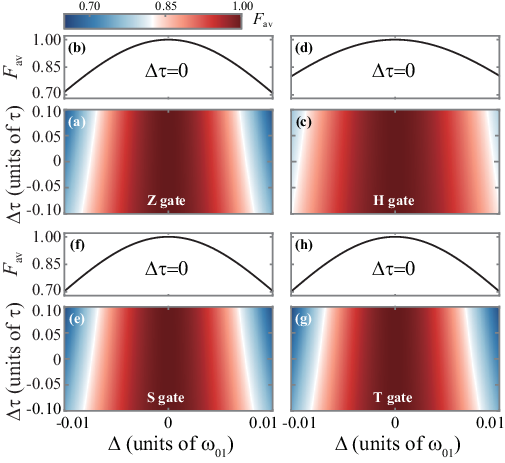}
    \caption{Average gate fidelity $F_\text{av}$ as a function of pulse detuning $\Delta$ and delay-time error $\Delta\tau$. Numerical simulations are performed for the (a, b) Pauli-Z, (c, d) Hadamard, (e, f) S, and (g, h) T gates. The lower panels (a, c, e, g) show heatmaps of $F_\text{av}$ versus $\Delta\tau$ and $\Delta$, where the detuning $\Delta$ and delay error $\Delta\tau$ are scaled in units of $\omega_{01}$ and $\tau$, respectively. The upper panels (b, d, f, h) present the corresponding line graphs of $F_\text{av}$ versus $\Delta$ extracted at the condition of zero delay time error ($\Delta\tau = 0$).
     }
    \label{fig3}
\end{figure}
\begin{figure*}[!ht]
    \centering        \includegraphics[width=0.9\textwidth]{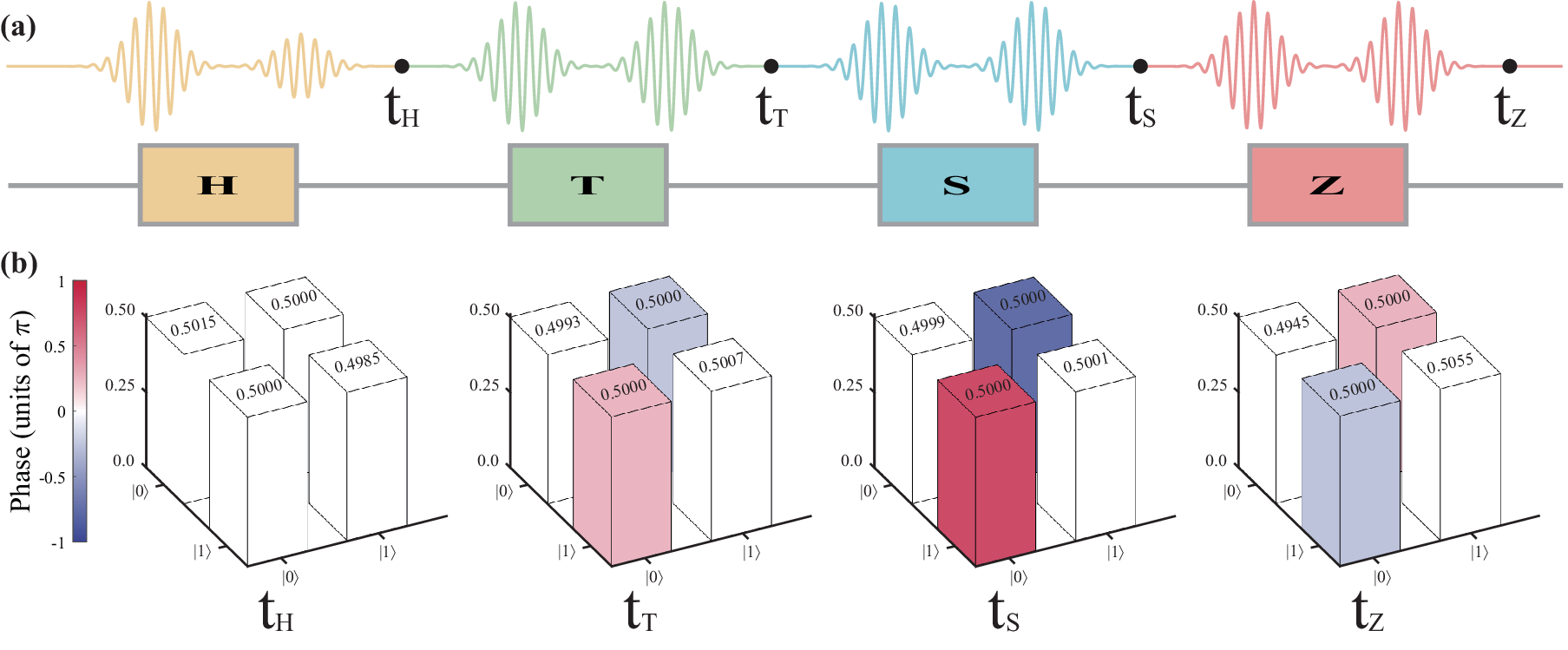}
    \caption{Demonstration of a sequential quantum circuit and state evolution. (a) Schematic of the quantum circuit (lower) and the corresponding time-domain pulse sequence (upper). The circuit initializes the molecule in the $|0\rangle$ state, followed by a Hadamard gate and a sequence of T, S, and Z gates. Pulse colors correspond to their respective logic gates, with $t_i$ ($i=\text{H, T, S, Z}$) denoting the end time of each operation. (b) Evolution of the density matrix $\rho$ at each stage of the circuit. The numbers above the bars represent the absolute values of the density matrix elements $|\rho_{ij}|$, while the colors indicate the phase. All parameters are consistent with the ideal pulse conditions in Fig.~\ref{fig3}.}
    \label{fig4}
\end{figure*}
We systematically assess the robustness of the proposed control scheme and the trade-off between gate speed and operational fidelity under realistic conditions by analyzing gate performance with pulse imperfections. We consider cases where both control pulses have identical bandwidth errors ($\Delta\omega$) and phase errors ($\Delta\phi$). The inter-pulse delay is set to $\tau = 112\tau_0$ (where $\tau_0 = \pi/B$ is the full revival time) to prevent coherent superposition when pulses share a common frequency. Figure~\ref{fig2} presents the average gate fidelity $F_\text{av}$ for the Pauli-Z, Hadamard, S, and T gates as a function of these error parameters. Table~\ref{tab:gates} provides detailed numerical settings.\\ \indent 
Figures~\ref{fig2}(a), (c), (e), and (g) show two-dimensional heatmaps of $F_\text{av}$ sensitivity across the $(\Delta\omega, \Delta\phi)$ parameter space. Figures~\ref{fig2}(b), (d), (f), and (h) display one-dimensional line cuts at zero phase error ($\Delta\phi = 0$), illustrating the specific impact of pulse bandwidth. For all four gates, $F_{\text{av}}$ decreases as the pulse bandwidth $\Delta\omega$ increases. The fidelity remains high for narrow bandwidths but drops sharply in the large-bandwidth regime. This reduction is due to population leakage and unwanted off-resonant transitions, as excessive bandwidth allows spectral components to couple to channels outside the two-level computational subspace. These results underscore the need for narrow-bandwidth pulses to suppress non-adiabatic leakage and maintain high fidelity.\\ \indent 
While all gates show similar sensitivity to bandwidth, their responses to phase errors $\Delta\phi$ differ, distinguishing phase gates (Pauli-Z, S, and T) from the Hadamard gate. The phase gates are highly resilient to phase fluctuations, as indicated by the extended high-fidelity regions in Figs.~\ref{fig2}(a), (e), and (g). This resilience results from the scheme’s intrinsic compensation of diagonal phase errors. As shown in Eq.~(\ref{UT1}), the net phase accumulation depends on $(\phi_2 + \Delta\phi) - (\phi_1 + \Delta\phi) = \phi_2 - \phi_1$, which cancels any common-mode phase error $\Delta\phi$. In contrast, the Hadamard gate in Fig.~\ref{fig2}(c) shows rapid fidelity loss as $\Delta\phi$ deviates from zero. This sensitivity arises because the Hadamard operation relies on off-diagonal transitions. While diagonal phase errors are compensated, off-diagonal errors persist throughout the gate evolution, making the Hadamard gate the most vulnerable in the proposed set.\\ \indent 
To further evaluate the practical viability of the phase-locked control protocol, we analyze its sensitivity to two common experimental imperfections: pulse frequency detuning ($\Delta$) and pulse delay error ($\Delta\tau$), with both pulses affected equally. For these simulations, we use a pulse bandwidth of $\Delta\omega = 0.1\omega_{01}$ and a nominal delay of $\tau = 11.2\tau_0$. These parameters, chosen based on the optimization landscape in Fig.~\ref{fig2}, balance fast gate operation with suppression of non-adiabatic leakage. Under these optimized conditions, the complete gate operation takes approximately 8 ns, which is seven orders of magnitude shorter than the 250 ms coherence time recently demonstrated for the NaCs lowest two rotational manifolds\cite{park2023extended}. This fast operation in time ensures that environmental decoherence is negligible during gate evolution. \\ \indent 
Figure~\ref{fig3} shows two-dimensional maps of $F_{\text{av}}$ for all four gates across the $(\Delta, \Delta\tau)$ parameter space in the lower panels [Figs.~\ref{fig3}(a), (c), (e), and (g)], and high-resolution line cuts of fidelity decay as a function of detuning $\Delta$ at zero delay error ($\Delta\tau = 0$) in the upper panels [Figs.~\ref{fig3}(b), (d), (f), and (h)]. The contour maps demonstrate strong resilience to time-delay errors. The average fidelity remains above $0.9999$ even when $\Delta\tau$ reaches 10\% of the nominal pulse interval. This robustness results from the phase-locking mechanism. Unlike conventional schemes, where the relative phase depends on arrival time, our phase-locked pulses maintain coherence across the carrier phase. Provided the pulses are temporally separated to avoid overlap, gate performance is largely unaffected by variations in arrival time.\\ \indent 
\begin{figure*}[htbp]
    \centering        \includegraphics[width=1.0\textwidth]{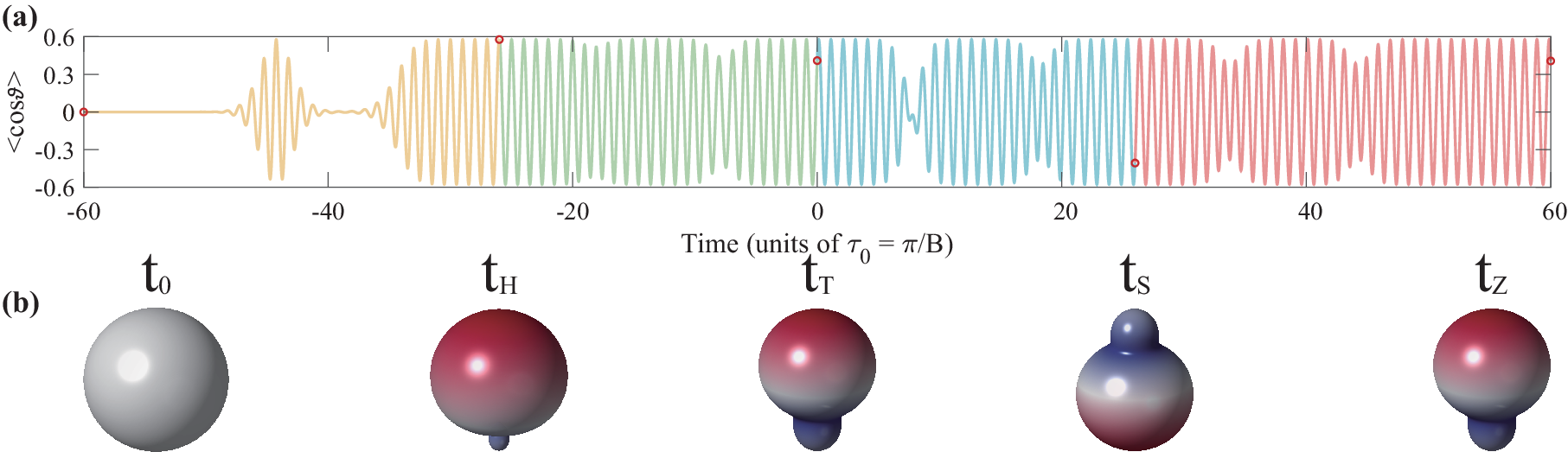}
    \caption{Rotational dynamics and wave packet evolution under the quantum circuit in Fig.~\ref{fig4}. (a) Time-dependent molecular orientation $\langle\cos\vartheta\rangle$. The color-coded lines correspond to the durations of the H, T, S, and Z gates as defined in Fig. 4(a). Red open circles mark the start and end of each gate operation. (b) Three-dimensional snapshots of the molecular rotational wave packets at the five marked timestamps in (a).
     }
    \label{fig5}
\end{figure*}
The trapezoidal shape of the high-fidelity regions highlights a key trade-off: robustness to detuning depends on the inter-pulse delay. Shorter intervals provide greater tolerance to detuning because, during idle periods, any detuning $\Delta$ acts as a residual longitudinal field and causes a phase error $\delta \phi = \Delta \cdot \tau_{\text{idle}}$. Reducing the pulse interval limits error accumulation and improves tolerance to detuning. The line graphs in the upper panels show the system's vulnerability to frequency deviations. Even with short delays, fidelity declines rapidly as $\Delta$ increases, and operational utility drops below the 1\% error threshold. There exists a clear distinction between gate types: phase gates are highly sensitive to detuning-induced longitudinal rotations, while the Hadamard gate is more tolerant. The Hadamard operation features significant off-diagonal couplings, which reduce phase errors through dynamical decoupling and make it the most robust against frequency fluctuations.
\subsection{Simulations for quantum circuits}
We evaluate the practical performance of the phase-locked control scheme by simulating the sequential execution of a composite quantum circuit. Pulse parameters are set to the optimal, high-fidelity point identified in Fig.~\ref{fig3}. Figure~\ref{fig4} displays the circuit schematic and corresponding real-time state evolution. In Fig.~\ref{fig4}(a), the system is initialized in the ground state $|0\rangle$. Since phase gates are diagonal and do not alter computational basis populations, a Hadamard gate is first applied to prepare the molecule in a maximally coherent superposition, $(|0\rangle + |1\rangle)/\sqrt{2}$. The T, S, and Z gates are then applied sequentially to verify precise phase accumulation. The upper panel shows the time-domain pulse sequence, with color coding and vertical alignment corresponding to the logic gates in the circuit schematic. Timestamps $t_{\text{H}}$, $t_{\text{T}}$, $t_{\text{S}}$, and $t_{\text{Z}}$ mark the completion of each operation.\\ \indent 
Figure~\ref{fig4}(b) shows the evolution of the system’s density matrix $\rho$. Each bar chart represents the quantum state after a given gate operation. Labels above the bars indicate magnitudes $|\rho_{ij}|$, and bar colors represent phase values $\arg(\rho_{ij})$. After the initial Hadamard gate, all density matrix element magnitudes remain near $0.5$, confirming a balanced superposition with negligible leakage into auxiliary states. The changing bar colors reflect the evolution of the relative phase driven by the sequential T, S, and Z gates. A sudden color “flip” between $t_{\text{S}}$ and $t_{\text{Z}}$ indicates phase wrapping, not a physical discontinuity. When the cumulative phase exceeds $+\pi$, it is mapped to the $[-\pi, \pi]$ interval (modulo $2\pi$) to match the color bar scale. The cumulative fidelity remains high at 0.9999 throughout. These results demonstrate the scalability and speed of our gating scheme, as well as its potential to support complex, multi-gate quantum logic operations in rotational molecular qubits.\\ \indent 
Because the molecular qubit comprises a superposition of rotational states $|0\rangle$ and $|1\rangle$, it forms a rotational wave packet that evolves freely once the control fields are extinguished. This field-free evolution manifests as periodic molecular orientation  \cite{koch2019quantum,mun2022all,hong2023quantum}. The degree of orientation, quantified by $\langle\cos\vartheta\rangle$, is typically measured using velocity map imaging\cite{tutunnikov2020observation}, ultrafast electron or X-ray diffraction\cite{yang2016diffractive}, coulomb explosion imaging\cite{trippel2015two}, or polarized spectroscopy\cite{renard2003postpulse}. For quantum information applications, weak-field polarization spectroscopy may offer a non-destructive alternative to ionization-based methods  \cite{renard2003postpulse,peng2015measurement,lian2023visualizing}. In this approach, a weak, linearly polarized probe laser passes through the molecular ensemble, and the resulting birefringence rotates the polarization plane by an angle proportional to $\langle\cos\vartheta\rangle$, the first moment of the angular distribution. This technique allows quantitative extraction of the orientation signal and provides real-time access to both the amplitude and phase of the rotational qubit.\\ \indent 
For a general superposition state $|\psi(t)\rangle = c_0(t)|0,0\rangle + c_1(t)|1,0\rangle$ (with $M=0$ selected by linear excitation), the expectation value evolves as
\begin{equation}
\langle\cos\vartheta\rangle(t) = A\cos(\omega_{01}t - \varphi_{01}),
\end{equation}
where the oscillation amplitude
\begin{equation}
A = \frac{2\sqrt{3}}{3}|c_0||c_1|
\end{equation}
reflects the degree of coherence between the two rotational states, and $\varphi_{01} = \arg(c_1) - \arg(c_0)$ denotes the relative quantum phase. The amplitude reaches its theoretical maximum $A_{\max} = \sqrt{3}/3 \approx 0.577$ for an equatorial superposition ($|c_0| = |c_1| = 1/\sqrt{2}$), vanishing for pure eigenstates. Crucially, any phase shift induced by quantum gate operations translates into a temporal displacement $\Delta t = \varphi_{01}/\omega_{01}$ of the oscillation extrema, enabling direct readout of gate phases without full state reconstruction. Monitoring these polarization dynamics thus permits complete characterization of pure-state qubit evolution and provides a direct diagnostic of single-qubit gate operations.\\ \indent
Figure~\ref{fig5}(a) shows the time-resolved evolution of $\langle\cos\vartheta\rangle$ for the gate sequence in Fig.~\ref{fig4}(a). Red hollow circles mark the initial state at $t_0$ and the pulse endpoints. Before the Hadamard gate ($t < t_{\text{H}}$), $\langle\cos\vartheta\rangle = 0$, confirming the system is in the pure ground state $|0\rangle$. After the Hadamard gate, oscillations appear with period $\tau_0$ and peak amplitude 0.577, confirming the maximal superposition. Subsequent phase gates preserve the amplitude but shift the extrema by $t’$, where $\omega_{01}t’ - \varphi_{01} = 0$. Measuring this shift determines the qubit phase $\varphi_{01}$. Figure \ref{fig5}(b) displays rotational wave packet distributions at selected times. The isotropic distribution at $t_0$ becomes anisotropic after the Hadamard gate, indicating coherent superposition. The sign of $\langle\cos\vartheta\rangle$ indicates orientation: positive values correspond to alignment along the quantization axis ($\vartheta < \pi/2$), while negative values correspond to alignment in the opposite direction ($\vartheta > \pi/2$).
\section{CONCLUSION AND OUTLOOK}\label{Sec:Conclusion}
We have developed and validated a theoretical framework for implementing universal single-qubit gates using rotational states of ultracold polar molecules. By encoding qubits in the lowest rotational levels of NaCs and employing an optimized two-pulse protocol derived from first-order Magnus expansion, we demonstrate arbitrary single-qubit rotations with fidelities exceeding $0.9999$ and negligible leakage into auxiliary states. Our robustness analysis highlights several key points. Phase gates, including Pauli-Z, S, and T, are more resilient to common-mode phase fluctuations than the Hadamard gate due to inherent phase-error compensation. All gates require narrow-bandwidth pulses to minimize leakage. The phase-locked scheme maintains a fidelity above $0.9999$ even when delay errors are ten times the nominal interval. There is a trade-off between detuning robustness and inter-pulse delay: shorter intervals improve frequency tolerance by reducing phase-error accumulation, and the Hadamard gate receives additional protection against off-diagonal couplings from dynamical decoupling.
Composite circuit simulations show that high fidelity is maintained during sequential multi-gate operations. The Hadamard gate consistently prepares balanced superpositions, whereas phase gates accumulate relative phases without affecting the populations in the computational basis. The molecular orientation degree $\langle\cos\vartheta\rangle(t)$, measured by weak-field polarization rotation, offers a direct experimental state readout.\\ \indent 
Regarding experimental feasibility, state-of-the-art techniques have achieved overall rovibronic ground-state preparation with over 31(4)\% efficiency \cite{cairncross2021assembly}, optical tweezer trapping with millisecond coherence times \cite{park2023extended}, and microwave-driven transitions with Rabi frequencies above 1 GHz \cite{yoshihara2014flux}. The approximately 378 MHz bandwidth requirement matches that of current, mature waveform generators \cite{bardin2021microwaves}, and the phase-locked sequence is consistent with established Ramsey interferometry \cite{cronin2009optics}. The large NaCs permanent dipole moment provides strong orientation signals for polarization readout. Integrating optical lattice clock technologies could further enhance coherence through magic-wavelength trapping \cite{borkowski2018optical}, while parallel tweezer addressing supports scalable multi-qubit architectures \cite{zhang2022optical,picard2025entanglement}. Furthermore, our analytical method is also applicable to other molecules, and therefore this work outlines a clear path toward high-fidelity, scalable molecular quantum information processors with accessible experimental characterization.\\ \indent 
\begin{acknowledgments}
This work was supported by the National Natural Science Foundation of China under Grant No. 12274470. 
The simulation was conducted using the computing resources of the High Performance Computing Center at Central South University.
    \end{acknowledgments}

\appendix
\section{Magic Angle Configuration for Elimination of Tensor Light Shifts}\label{appendix}
In experiments involving optically trapped polar molecules, the AC Stark interaction with the trapping laser introduces state-dependent energy shifts. The shift of a specific sublevel can be decomposed into scalar and tensor contributions: $\Delta E = \Delta E_{\text{scalar}} + \Delta E_{\text{tensor}}$. While the scalar term shifts all magnetic sublevels equally, the tensor term lifts the degeneracy of the $m_F$ states. This differential shift causes inhomogeneous dephasing and limits the coherence time for rotational or hyperfine qubits.

To eliminate the tensor contribution, we align the polarization of the linearly polarized trapping field at the "magic angle" with respect to the quantization axis. As illustrated in Fig.\ref{figure1}(a), the uniform magnetic bias field $\vec{B}$ defines the quantization $z$ axis for the NaCs molecules.  $\theta'$ denotes the angle between the trapping electric field vector $\vec{\varepsilon}_T$ and the external magnetic field $\vec{B}$. The tensor term scales with the second-order Legendre polynomial \cite{neyenhuis2012anisotropic,zhang2024dressed}:

$$ \Delta E_{\text{tensor}} \propto P_2(\cos\theta') = \frac{1}{2}(3\cos^2\theta' - 1). $$

The tensor shift vanishes when the geometric pre-factor equals zero. This defines the magic angle condition: $3\cos^2\theta_{\text{magic}} - 1 = 0$, and therefore leads to the magic angle 

$$ \theta_{\text{magic}} = \arccos\left(\frac{1}{\sqrt{3}}\right) \approx 54.74^\circ. $$

 The optical dipole trap is prepared in a pure linear polarization state using a combination of polarizing optics. A half-wave plate mounted on a precision rotation stage is used to adjust the polarization direction $\vec{\varepsilon}_T$  such that $\theta' = \theta_{\text{magic}}$ relative to $\vec{B}$. Experiments have verified the magic angle configuration by performing microwave spectroscopy on the rotational states \cite{neyenhuis2012anisotropic,seesselberg2018extending,burchesky2021rotational}. At the magic angle, the differential light shift between these sublevels is minimized, confirming the suppression of the tensor light shift.


\begin{thebibliography}{72}%
\makeatletter
\providecommand \@ifxundefined [1]{%
 \@ifx{#1\undefined}
}%
\providecommand \@ifnum [1]{%
 \ifnum #1\expandafter \@firstoftwo
 \else \expandafter \@secondoftwo
 \fi
}%
\providecommand \@ifx [1]{%
 \ifx #1\expandafter \@firstoftwo
 \else \expandafter \@secondoftwo
 \fi
}%
\providecommand \natexlab [1]{#1}%
\providecommand \enquote  [1]{``#1''}%
\providecommand \bibnamefont  [1]{#1}%
\providecommand \bibfnamefont [1]{#1}%
\providecommand \citenamefont [1]{#1}%
\providecommand \href@noop [0]{\@secondoftwo}%
\providecommand \href [0]{\begingroup \@sanitize@url \@href}%
\providecommand \@href[1]{\@@startlink{#1}\@@href}%
\providecommand \@@href[1]{\endgroup#1\@@endlink}%
\providecommand \@sanitize@url [0]{\catcode `\\12\catcode `\$12\catcode `\&12\catcode `\#12\catcode `\^12\catcode `\_12\catcode `\%12\relax}%
\providecommand \@@startlink[1]{}%
\providecommand \@@endlink[0]{}%
\providecommand \url  [0]{\begingroup\@sanitize@url \@url }%
\providecommand \@url [1]{\endgroup\@href {#1}{\urlprefix }}%
\providecommand \urlprefix  [0]{URL }%
\providecommand \Eprint [0]{\href }%
\providecommand \doibase [0]{https://doi.org/}%
\providecommand \selectlanguage [0]{\@gobble}%
\providecommand \bibinfo  [0]{\@secondoftwo}%
\providecommand \bibfield  [0]{\@secondoftwo}%
\providecommand \translation [1]{[#1]}%
\providecommand \BibitemOpen [0]{}%
\providecommand \bibitemStop [0]{}%
\providecommand \bibitemNoStop [0]{.\EOS\space}%
\providecommand \EOS [0]{\spacefactor3000\relax}%
\providecommand \BibitemShut  [1]{\csname bibitem#1\endcsname}%
\let\auto@bib@innerbib\@empty
\bibitem [{\citenamefont {Steane}(1998)}]{steane1998quantum}%
  \BibitemOpen
  \bibfield  {author} {\bibinfo {author} {\bibfnamefont {A.}~\bibnamefont {Steane}},\ }\bibfield  {title} {\bibinfo {title} {Quantum computing},\ }\href {https://doi.org/10.1088/0034-4885/61/2/002} {\bibfield  {journal} {\bibinfo  {journal} {Rep. Prog. Phys.}\ }\textbf {\bibinfo {volume} {61}},\ \bibinfo {pages} {117} (\bibinfo {year} {1998})}\BibitemShut {NoStop}%
\bibitem [{\citenamefont {Preskill}(1998)}]{preskill1998reliable}%
  \BibitemOpen
  \bibfield  {author} {\bibinfo {author} {\bibfnamefont {J.}~\bibnamefont {Preskill}},\ }\bibfield  {title} {\bibinfo {title} {Reliable quantum computers},\ }\href {https://doi.org/https://doi.org/10.1098/rspa.1998.0167} {\bibfield  {journal} {\bibinfo  {journal} {Proc. R. Soc. Lond. A}\ }\textbf {\bibinfo {volume} {454}},\ \bibinfo {pages} {385} (\bibinfo {year} {1998})}\BibitemShut {NoStop}%
\bibitem [{\citenamefont {Ladd}\ \emph {et~al.}(2010)\citenamefont {Ladd}, \citenamefont {Jelezko}, \citenamefont {Laflamme}, \citenamefont {Nakamura}, \citenamefont {Monroe},\ and\ \citenamefont {O’Brien}}]{ladd2010quantum}%
  \BibitemOpen
  \bibfield  {author} {\bibinfo {author} {\bibfnamefont {T.~D.}\ \bibnamefont {Ladd}}, \bibinfo {author} {\bibfnamefont {F.}~\bibnamefont {Jelezko}}, \bibinfo {author} {\bibfnamefont {R.}~\bibnamefont {Laflamme}}, \bibinfo {author} {\bibfnamefont {Y.}~\bibnamefont {Nakamura}}, \bibinfo {author} {\bibfnamefont {C.}~\bibnamefont {Monroe}},\ and\ \bibinfo {author} {\bibfnamefont {J.~L.}\ \bibnamefont {O’Brien}},\ }\bibfield  {title} {\bibinfo {title} {Quantum computers},\ }\href {https://doi.org/https://doi.org/10.1038/nature08812} {\bibfield  {journal} {\bibinfo  {journal} {Nature}\ }\textbf {\bibinfo {volume} {464}},\ \bibinfo {pages} {45} (\bibinfo {year} {2010})}\BibitemShut {NoStop}%
\bibitem [{\citenamefont {Campbell}\ \emph {et~al.}(2017)\citenamefont {Campbell}, \citenamefont {Terhal},\ and\ \citenamefont {Vuillot}}]{campbell2017roads}%
  \BibitemOpen
  \bibfield  {author} {\bibinfo {author} {\bibfnamefont {E.~T.}\ \bibnamefont {Campbell}}, \bibinfo {author} {\bibfnamefont {B.~M.}\ \bibnamefont {Terhal}},\ and\ \bibinfo {author} {\bibfnamefont {C.}~\bibnamefont {Vuillot}},\ }\bibfield  {title} {\bibinfo {title} {Roads towards fault-tolerant universal quantum computation},\ }\href {https://doi.org/https://doi.org/10.1038/nature23460} {\bibfield  {journal} {\bibinfo  {journal} {Nature}\ }\textbf {\bibinfo {volume} {549}},\ \bibinfo {pages} {172} (\bibinfo {year} {2017})}\BibitemShut {NoStop}%
\bibitem [{\citenamefont {Zhou}\ \emph {et~al.}(2020)\citenamefont {Zhou}, \citenamefont {Stoudenmire},\ and\ \citenamefont {Waintal}}]{zhou2020limits}%
  \BibitemOpen
  \bibfield  {author} {\bibinfo {author} {\bibfnamefont {Y.}~\bibnamefont {Zhou}}, \bibinfo {author} {\bibfnamefont {E.~M.}\ \bibnamefont {Stoudenmire}},\ and\ \bibinfo {author} {\bibfnamefont {X.}~\bibnamefont {Waintal}},\ }\bibfield  {title} {\bibinfo {title} {What limits the simulation of quantum computers?},\ }\href {https://doi.org/https://doi.org/10.1103/PhysRevX.10.041038} {\bibfield  {journal} {\bibinfo  {journal} {Phys. Rev. X}\ }\textbf {\bibinfo {volume} {10}},\ \bibinfo {pages} {041038} (\bibinfo {year} {2020})}\BibitemShut {NoStop}%
\bibitem [{\citenamefont {Wendin}(2017)}]{wendin2017quantum}%
  \BibitemOpen
  \bibfield  {author} {\bibinfo {author} {\bibfnamefont {G.}~\bibnamefont {Wendin}},\ }\bibfield  {title} {\bibinfo {title} {Quantum information processing with superconducting circuits: a review},\ }\href {https://doi.org/10.1088/1361-6633/aa7e1a} {\bibfield  {journal} {\bibinfo  {journal} {Rep. Prog. Phys.}\ }\textbf {\bibinfo {volume} {80}},\ \bibinfo {pages} {106001} (\bibinfo {year} {2017})}\BibitemShut {NoStop}%
\bibitem [{\citenamefont {Gu}\ \emph {et~al.}(2017)\citenamefont {Gu}, \citenamefont {Kockum}, \citenamefont {Miranowicz}, \citenamefont {Liu},\ and\ \citenamefont {Nori}}]{gu2017microwave}%
  \BibitemOpen
  \bibfield  {author} {\bibinfo {author} {\bibfnamefont {X.}~\bibnamefont {Gu}}, \bibinfo {author} {\bibfnamefont {A.~F.}\ \bibnamefont {Kockum}}, \bibinfo {author} {\bibfnamefont {A.}~\bibnamefont {Miranowicz}}, \bibinfo {author} {\bibfnamefont {Y.-X.}\ \bibnamefont {Liu}},\ and\ \bibinfo {author} {\bibfnamefont {F.}~\bibnamefont {Nori}},\ }\bibfield  {title} {\bibinfo {title} {Microwave photonics with superconducting quantum circuits},\ }\href {https://doi.org/https://doi.org/10.1016/j.physrep.2017.10.002} {\bibfield  {journal} {\bibinfo  {journal} {Phys. Rep.}\ }\textbf {\bibinfo {volume} {718}},\ \bibinfo {pages} {1} (\bibinfo {year} {2017})}\BibitemShut {NoStop}%
\bibitem [{\citenamefont {Krantz}\ \emph {et~al.}(2019)\citenamefont {Krantz}, \citenamefont {Kjaergaard}, \citenamefont {Yan}, \citenamefont {Orlando}, \citenamefont {Gustavsson},\ and\ \citenamefont {Oliver}}]{krantz2019quantum}%
  \BibitemOpen
  \bibfield  {author} {\bibinfo {author} {\bibfnamefont {P.}~\bibnamefont {Krantz}}, \bibinfo {author} {\bibfnamefont {M.}~\bibnamefont {Kjaergaard}}, \bibinfo {author} {\bibfnamefont {F.}~\bibnamefont {Yan}}, \bibinfo {author} {\bibfnamefont {T.~P.}\ \bibnamefont {Orlando}}, \bibinfo {author} {\bibfnamefont {S.}~\bibnamefont {Gustavsson}},\ and\ \bibinfo {author} {\bibfnamefont {W.~D.}\ \bibnamefont {Oliver}},\ }\bibfield  {title} {\bibinfo {title} {A quantum engineer's guide to superconducting qubits},\ }\href {https://doi.org/https://doi.org/10.1063/1.5089550} {\bibfield  {journal} {\bibinfo  {journal} {Appl. Phys. Rev.}\ }\textbf {\bibinfo {volume} {6}},\ \bibinfo {pages} {021318} (\bibinfo {year} {2019})}\BibitemShut {NoStop}%
\bibitem [{\citenamefont {Siddiqi}(2021)}]{siddiqi2021engineering}%
  \BibitemOpen
  \bibfield  {author} {\bibinfo {author} {\bibfnamefont {I.}~\bibnamefont {Siddiqi}},\ }\bibfield  {title} {\bibinfo {title} {Engineering high-coherence superconducting qubits},\ }\href {https://doi.org/https://doi.org/10.1038/s41578-021-00370-4} {\bibfield  {journal} {\bibinfo  {journal} {Nat. Rev. Mater.}\ }\textbf {\bibinfo {volume} {6}},\ \bibinfo {pages} {875} (\bibinfo {year} {2021})}\BibitemShut {NoStop}%
\bibitem [{\citenamefont {Gyenis}\ \emph {et~al.}(2021)\citenamefont {Gyenis}, \citenamefont {Mundada}, \citenamefont {Di~Paolo}, \citenamefont {Hazard}, \citenamefont {You}, \citenamefont {Schuster}, \citenamefont {Koch}, \citenamefont {Blais},\ and\ \citenamefont {Houck}}]{gyenis2021experimental}%
  \BibitemOpen
  \bibfield  {author} {\bibinfo {author} {\bibfnamefont {A.}~\bibnamefont {Gyenis}}, \bibinfo {author} {\bibfnamefont {P.~S.}\ \bibnamefont {Mundada}}, \bibinfo {author} {\bibfnamefont {A.}~\bibnamefont {Di~Paolo}}, \bibinfo {author} {\bibfnamefont {T.~M.}\ \bibnamefont {Hazard}}, \bibinfo {author} {\bibfnamefont {X.}~\bibnamefont {You}}, \bibinfo {author} {\bibfnamefont {D.~I.}\ \bibnamefont {Schuster}}, \bibinfo {author} {\bibfnamefont {J.}~\bibnamefont {Koch}}, \bibinfo {author} {\bibfnamefont {A.}~\bibnamefont {Blais}},\ and\ \bibinfo {author} {\bibfnamefont {A.~A.}\ \bibnamefont {Houck}},\ }\bibfield  {title} {\bibinfo {title} {Experimental realization of a protected superconducting circuit derived from the 0--$\pi$ qubit},\ }\href {https://doi.org/https://doi.org/10.1103/PRXQuantum.2.010339} {\bibfield  {journal} {\bibinfo  {journal} {PRX Quantum}\ }\textbf {\bibinfo {volume} {2}},\ \bibinfo {pages} {010339} (\bibinfo {year} {2021})}\BibitemShut {NoStop}%
\bibitem [{\citenamefont {Ganjam}\ \emph {et~al.}(2024)\citenamefont {Ganjam}, \citenamefont {Wang}, \citenamefont {Lu}, \citenamefont {Banerjee}, \citenamefont {Lei}, \citenamefont {Krayzman}, \citenamefont {Kisslinger}, \citenamefont {Zhou}, \citenamefont {Li}, \citenamefont {Jia} \emph {et~al.}}]{ganjam2024surpassing}%
  \BibitemOpen
  \bibfield  {author} {\bibinfo {author} {\bibfnamefont {S.}~\bibnamefont {Ganjam}}, \bibinfo {author} {\bibfnamefont {Y.}~\bibnamefont {Wang}}, \bibinfo {author} {\bibfnamefont {Y.}~\bibnamefont {Lu}}, \bibinfo {author} {\bibfnamefont {A.}~\bibnamefont {Banerjee}}, \bibinfo {author} {\bibfnamefont {C.~U.}\ \bibnamefont {Lei}}, \bibinfo {author} {\bibfnamefont {L.}~\bibnamefont {Krayzman}}, \bibinfo {author} {\bibfnamefont {K.}~\bibnamefont {Kisslinger}}, \bibinfo {author} {\bibfnamefont {C.}~\bibnamefont {Zhou}}, \bibinfo {author} {\bibfnamefont {R.}~\bibnamefont {Li}}, \bibinfo {author} {\bibfnamefont {Y.}~\bibnamefont {Jia}}, \emph {et~al.},\ }\bibfield  {title} {\bibinfo {title} {Surpassing millisecond coherence in on chip superconducting quantum memories by optimizing materials and circuit design},\ }\href {https://doi.org/https://doi.org/10.1038/s41467-024-47857-6} {\bibfield  {journal} {\bibinfo  {journal} {Nat. Commun.}\ }\textbf {\bibinfo {volume} {15}},\ \bibinfo {pages} {3687} (\bibinfo {year}
  {2024})}\BibitemShut {NoStop}%
\bibitem [{\citenamefont {Maurand}\ \emph {et~al.}(2016)\citenamefont {Maurand}, \citenamefont {Jehl}, \citenamefont {Kotekar-Patil}, \citenamefont {Corna}, \citenamefont {Bohuslavskyi}, \citenamefont {Lavi{\'e}ville}, \citenamefont {Hutin}, \citenamefont {Barraud}, \citenamefont {Vinet}, \citenamefont {Sanquer} \emph {et~al.}}]{maurand2016cmos}%
  \BibitemOpen
  \bibfield  {author} {\bibinfo {author} {\bibfnamefont {R.}~\bibnamefont {Maurand}}, \bibinfo {author} {\bibfnamefont {X.}~\bibnamefont {Jehl}}, \bibinfo {author} {\bibfnamefont {D.}~\bibnamefont {Kotekar-Patil}}, \bibinfo {author} {\bibfnamefont {A.}~\bibnamefont {Corna}}, \bibinfo {author} {\bibfnamefont {H.}~\bibnamefont {Bohuslavskyi}}, \bibinfo {author} {\bibfnamefont {R.}~\bibnamefont {Lavi{\'e}ville}}, \bibinfo {author} {\bibfnamefont {L.}~\bibnamefont {Hutin}}, \bibinfo {author} {\bibfnamefont {S.}~\bibnamefont {Barraud}}, \bibinfo {author} {\bibfnamefont {M.}~\bibnamefont {Vinet}}, \bibinfo {author} {\bibfnamefont {M.}~\bibnamefont {Sanquer}}, \emph {et~al.},\ }\bibfield  {title} {\bibinfo {title} {A {CMOS} silicon spin qubit},\ }\href {https://doi.org/https://doi.org/10.1038/ncomms13575} {\bibfield  {journal} {\bibinfo  {journal} {Nat. Commun.}\ }\textbf {\bibinfo {volume} {7}},\ \bibinfo {pages} {13575} (\bibinfo {year} {2016})}\BibitemShut {NoStop}%
\bibitem [{\citenamefont {Veldhorst}\ \emph {et~al.}(2017)\citenamefont {Veldhorst}, \citenamefont {Eenink}, \citenamefont {Yang},\ and\ \citenamefont {Dzurak}}]{veldhorst2017silicon}%
  \BibitemOpen
  \bibfield  {author} {\bibinfo {author} {\bibfnamefont {M.}~\bibnamefont {Veldhorst}}, \bibinfo {author} {\bibfnamefont {H.~G.}\ \bibnamefont {Eenink}}, \bibinfo {author} {\bibfnamefont {C.-H.}\ \bibnamefont {Yang}},\ and\ \bibinfo {author} {\bibfnamefont {A.~S.}\ \bibnamefont {Dzurak}},\ }\bibfield  {title} {\bibinfo {title} {Silicon {CMOS} architecture for a spin-based quantum computer},\ }\href {https://doi.org/https://doi.org/10.1038/s41467-017-01905-6} {\bibfield  {journal} {\bibinfo  {journal} {Nat. Commun.}\ }\textbf {\bibinfo {volume} {8}},\ \bibinfo {pages} {1766} (\bibinfo {year} {2017})}\BibitemShut {NoStop}%
\bibitem [{\citenamefont {Gonzalez-Zalba}\ \emph {et~al.}(2021)\citenamefont {Gonzalez-Zalba}, \citenamefont {De~Franceschi}, \citenamefont {Charbon}, \citenamefont {Meunier}, \citenamefont {Vinet},\ and\ \citenamefont {Dzurak}}]{gonzalez2021scaling}%
  \BibitemOpen
  \bibfield  {author} {\bibinfo {author} {\bibfnamefont {M.}~\bibnamefont {Gonzalez-Zalba}}, \bibinfo {author} {\bibfnamefont {S.}~\bibnamefont {De~Franceschi}}, \bibinfo {author} {\bibfnamefont {E.}~\bibnamefont {Charbon}}, \bibinfo {author} {\bibfnamefont {T.}~\bibnamefont {Meunier}}, \bibinfo {author} {\bibfnamefont {M.}~\bibnamefont {Vinet}},\ and\ \bibinfo {author} {\bibfnamefont {A.}~\bibnamefont {Dzurak}},\ }\bibfield  {title} {\bibinfo {title} {Scaling silicon-based quantum computing using {CMOS} technology},\ }\href {https://doi.org/https://doi.org/10.1038/s41928-021-00681-y} {\bibfield  {journal} {\bibinfo  {journal} {Nat. Electron.}\ }\textbf {\bibinfo {volume} {4}},\ \bibinfo {pages} {872} (\bibinfo {year} {2021})}\BibitemShut {NoStop}%
\bibitem [{\citenamefont {Yoneda}\ \emph {et~al.}(2018)\citenamefont {Yoneda}, \citenamefont {Takeda}, \citenamefont {Otsuka}, \citenamefont {Nakajima}, \citenamefont {Delbecq}, \citenamefont {Allison}, \citenamefont {Honda}, \citenamefont {Kodera}, \citenamefont {Oda}, \citenamefont {Hoshi} \emph {et~al.}}]{yoneda2018quantum}%
  \BibitemOpen
  \bibfield  {author} {\bibinfo {author} {\bibfnamefont {J.}~\bibnamefont {Yoneda}}, \bibinfo {author} {\bibfnamefont {K.}~\bibnamefont {Takeda}}, \bibinfo {author} {\bibfnamefont {T.}~\bibnamefont {Otsuka}}, \bibinfo {author} {\bibfnamefont {T.}~\bibnamefont {Nakajima}}, \bibinfo {author} {\bibfnamefont {M.~R.}\ \bibnamefont {Delbecq}}, \bibinfo {author} {\bibfnamefont {G.}~\bibnamefont {Allison}}, \bibinfo {author} {\bibfnamefont {T.}~\bibnamefont {Honda}}, \bibinfo {author} {\bibfnamefont {T.}~\bibnamefont {Kodera}}, \bibinfo {author} {\bibfnamefont {S.}~\bibnamefont {Oda}}, \bibinfo {author} {\bibfnamefont {Y.}~\bibnamefont {Hoshi}}, \emph {et~al.},\ }\bibfield  {title} {\bibinfo {title} {A quantum-dot spin qubit with coherence limited by charge noise and fidelity higher than 99.9\%},\ }\href {https://doi.org/https://doi.org/10.1038/s41565-017-0014-x} {\bibfield  {journal} {\bibinfo  {journal} {Nat. Nanotechnol.}\ }\textbf {\bibinfo {volume} {13}},\ \bibinfo {pages} {102} (\bibinfo {year}
  {2018})}\BibitemShut {NoStop}%
\bibitem [{\citenamefont {Huang}\ \emph {et~al.}(2018)\citenamefont {Huang}, \citenamefont {Zimmerman},\ and\ \citenamefont {Bryant}}]{huang2018spin}%
  \BibitemOpen
  \bibfield  {author} {\bibinfo {author} {\bibfnamefont {P.}~\bibnamefont {Huang}}, \bibinfo {author} {\bibfnamefont {N.~M.}\ \bibnamefont {Zimmerman}},\ and\ \bibinfo {author} {\bibfnamefont {G.~W.}\ \bibnamefont {Bryant}},\ }\bibfield  {title} {\bibinfo {title} {Spin decoherence in a two-qubit {CPHASE} gate: the critical role of tunneling noise},\ }\href {https://doi.org/https://doi.org/10.1038/s41534-018-0112-0} {\bibfield  {journal} {\bibinfo  {journal} {npj Quantum Inf.}\ }\textbf {\bibinfo {volume} {4}},\ \bibinfo {pages} {62} (\bibinfo {year} {2018})}\BibitemShut {NoStop}%
\bibitem [{\citenamefont {Shehata}\ \emph {et~al.}(2023)\citenamefont {Shehata}, \citenamefont {Simion}, \citenamefont {Li}, \citenamefont {Mohiyaddin}, \citenamefont {Wan}, \citenamefont {Mongillo}, \citenamefont {Govoreanu}, \citenamefont {Radu}, \citenamefont {De~Greve},\ and\ \citenamefont {Van~Dorpe}}]{shehata2023modeling}%
  \BibitemOpen
  \bibfield  {author} {\bibinfo {author} {\bibfnamefont {M.~M. E.~K.}\ \bibnamefont {Shehata}}, \bibinfo {author} {\bibfnamefont {G.}~\bibnamefont {Simion}}, \bibinfo {author} {\bibfnamefont {R.}~\bibnamefont {Li}}, \bibinfo {author} {\bibfnamefont {F.~A.}\ \bibnamefont {Mohiyaddin}}, \bibinfo {author} {\bibfnamefont {D.}~\bibnamefont {Wan}}, \bibinfo {author} {\bibfnamefont {M.}~\bibnamefont {Mongillo}}, \bibinfo {author} {\bibfnamefont {B.}~\bibnamefont {Govoreanu}}, \bibinfo {author} {\bibfnamefont {I.}~\bibnamefont {Radu}}, \bibinfo {author} {\bibfnamefont {K.}~\bibnamefont {De~Greve}},\ and\ \bibinfo {author} {\bibfnamefont {P.}~\bibnamefont {Van~Dorpe}},\ }\bibfield  {title} {\bibinfo {title} {Modeling semiconductor spin qubits and their charge noise environment for quantum gate fidelity estimation},\ }\href {https://doi.org/https://doi.org/10.1103/PhysRevB.108.045305} {\bibfield  {journal} {\bibinfo  {journal} {Phys. Rev. B}\ }\textbf {\bibinfo {volume} {108}},\ \bibinfo {pages} {045305} (\bibinfo
  {year} {2023})}\BibitemShut {NoStop}%
\bibitem [{\citenamefont {Harty}\ \emph {et~al.}(2014)\citenamefont {Harty}, \citenamefont {Allcock}, \citenamefont {Ballance}, \citenamefont {Guidoni}, \citenamefont {Janacek}, \citenamefont {Linke}, \citenamefont {Stacey},\ and\ \citenamefont {Lucas}}]{harty2014high}%
  \BibitemOpen
  \bibfield  {author} {\bibinfo {author} {\bibfnamefont {T.}~\bibnamefont {Harty}}, \bibinfo {author} {\bibfnamefont {D.}~\bibnamefont {Allcock}}, \bibinfo {author} {\bibfnamefont {C.~J.}\ \bibnamefont {Ballance}}, \bibinfo {author} {\bibfnamefont {L.}~\bibnamefont {Guidoni}}, \bibinfo {author} {\bibfnamefont {H.}~\bibnamefont {Janacek}}, \bibinfo {author} {\bibfnamefont {N.}~\bibnamefont {Linke}}, \bibinfo {author} {\bibfnamefont {D.}~\bibnamefont {Stacey}},\ and\ \bibinfo {author} {\bibfnamefont {D.}~\bibnamefont {Lucas}},\ }\bibfield  {title} {\bibinfo {title} {High-fidelity preparation, gates, memory, and readout of a trapped-ion quantum bit},\ }\href {https://doi.org/https://doi.org/10.1103/PhysRevLett.113.220501} {\bibfield  {journal} {\bibinfo  {journal} {Phys. Rev. Lett.}\ }\textbf {\bibinfo {volume} {113}},\ \bibinfo {pages} {220501} (\bibinfo {year} {2014})}\BibitemShut {NoStop}%
\bibitem [{\citenamefont {Wang}\ \emph {et~al.}(2017)\citenamefont {Wang}, \citenamefont {Um}, \citenamefont {Zhang}, \citenamefont {An}, \citenamefont {Lyu}, \citenamefont {Zhang}, \citenamefont {Duan}, \citenamefont {Yum},\ and\ \citenamefont {Kim}}]{wang2017single}%
  \BibitemOpen
  \bibfield  {author} {\bibinfo {author} {\bibfnamefont {Y.}~\bibnamefont {Wang}}, \bibinfo {author} {\bibfnamefont {M.}~\bibnamefont {Um}}, \bibinfo {author} {\bibfnamefont {J.}~\bibnamefont {Zhang}}, \bibinfo {author} {\bibfnamefont {S.}~\bibnamefont {An}}, \bibinfo {author} {\bibfnamefont {M.}~\bibnamefont {Lyu}}, \bibinfo {author} {\bibfnamefont {J.-N.}\ \bibnamefont {Zhang}}, \bibinfo {author} {\bibfnamefont {L.-M.}\ \bibnamefont {Duan}}, \bibinfo {author} {\bibfnamefont {D.}~\bibnamefont {Yum}},\ and\ \bibinfo {author} {\bibfnamefont {K.}~\bibnamefont {Kim}},\ }\bibfield  {title} {\bibinfo {title} {Single-qubit quantum memory exceeding ten-minute coherence time},\ }\href {https://doi.org/https://doi.org/10.1038/s41566-017-0007-1} {\bibfield  {journal} {\bibinfo  {journal} {Nat. Photon.}\ }\textbf {\bibinfo {volume} {11}},\ \bibinfo {pages} {646} (\bibinfo {year} {2017})}\BibitemShut {NoStop}%
\bibitem [{\citenamefont {Wang}\ \emph {et~al.}(2021)\citenamefont {Wang}, \citenamefont {Luan}, \citenamefont {Qiao}, \citenamefont {Um}, \citenamefont {Zhang}, \citenamefont {Wang}, \citenamefont {Yuan}, \citenamefont {Gu}, \citenamefont {Zhang},\ and\ \citenamefont {Kim}}]{wang2021single}%
  \BibitemOpen
  \bibfield  {author} {\bibinfo {author} {\bibfnamefont {P.}~\bibnamefont {Wang}}, \bibinfo {author} {\bibfnamefont {C.-Y.}\ \bibnamefont {Luan}}, \bibinfo {author} {\bibfnamefont {M.}~\bibnamefont {Qiao}}, \bibinfo {author} {\bibfnamefont {M.}~\bibnamefont {Um}}, \bibinfo {author} {\bibfnamefont {J.}~\bibnamefont {Zhang}}, \bibinfo {author} {\bibfnamefont {Y.}~\bibnamefont {Wang}}, \bibinfo {author} {\bibfnamefont {X.}~\bibnamefont {Yuan}}, \bibinfo {author} {\bibfnamefont {M.}~\bibnamefont {Gu}}, \bibinfo {author} {\bibfnamefont {J.}~\bibnamefont {Zhang}},\ and\ \bibinfo {author} {\bibfnamefont {K.}~\bibnamefont {Kim}},\ }\bibfield  {title} {\bibinfo {title} {Single ion qubit with estimated coherence time exceeding one hour},\ }\href {https://doi.org/https://doi.org/10.1038/s41467-020-20330-w} {\bibfield  {journal} {\bibinfo  {journal} {Nat. Commun.}\ }\textbf {\bibinfo {volume} {12}},\ \bibinfo {pages} {233} (\bibinfo {year} {2021})}\BibitemShut {NoStop}%
\bibitem [{\citenamefont {Monroe}\ and\ \citenamefont {Kim}(2013)}]{monroe2013scaling}%
  \BibitemOpen
  \bibfield  {author} {\bibinfo {author} {\bibfnamefont {C.}~\bibnamefont {Monroe}}\ and\ \bibinfo {author} {\bibfnamefont {J.}~\bibnamefont {Kim}},\ }\bibfield  {title} {\bibinfo {title} {Scaling the ion trap quantum processor},\ }\href {https://doi.org/https://doi.org/10.1126/science.1231298} {\bibfield  {journal} {\bibinfo  {journal} {Science}\ }\textbf {\bibinfo {volume} {339}},\ \bibinfo {pages} {1164} (\bibinfo {year} {2013})}\BibitemShut {NoStop}%
\bibitem [{\citenamefont {Friis}\ \emph {et~al.}(2018)\citenamefont {Friis}, \citenamefont {Marty}, \citenamefont {Maier}, \citenamefont {Hempel}, \citenamefont {Holz{\"a}pfel}, \citenamefont {Jurcevic}, \citenamefont {Plenio}, \citenamefont {Huber}, \citenamefont {Roos}, \citenamefont {Blatt} \emph {et~al.}}]{friis2018observation}%
  \BibitemOpen
  \bibfield  {author} {\bibinfo {author} {\bibfnamefont {N.}~\bibnamefont {Friis}}, \bibinfo {author} {\bibfnamefont {O.}~\bibnamefont {Marty}}, \bibinfo {author} {\bibfnamefont {C.}~\bibnamefont {Maier}}, \bibinfo {author} {\bibfnamefont {C.}~\bibnamefont {Hempel}}, \bibinfo {author} {\bibfnamefont {M.}~\bibnamefont {Holz{\"a}pfel}}, \bibinfo {author} {\bibfnamefont {P.}~\bibnamefont {Jurcevic}}, \bibinfo {author} {\bibfnamefont {M.~B.}\ \bibnamefont {Plenio}}, \bibinfo {author} {\bibfnamefont {M.}~\bibnamefont {Huber}}, \bibinfo {author} {\bibfnamefont {C.}~\bibnamefont {Roos}}, \bibinfo {author} {\bibfnamefont {R.}~\bibnamefont {Blatt}}, \emph {et~al.},\ }\bibfield  {title} {\bibinfo {title} {Observation of entangled states of a fully controlled 20-qubit system},\ }\href {https://doi.org/https://doi.org/10.1103/PhysRevX.8.021012} {\bibfield  {journal} {\bibinfo  {journal} {Phys. Rev. X}\ }\textbf {\bibinfo {volume} {8}},\ \bibinfo {pages} {021012} (\bibinfo {year} {2018})}\BibitemShut {NoStop}%
\bibitem [{\citenamefont {Kwon}\ \emph {et~al.}(2024)\citenamefont {Kwon}, \citenamefont {Setzer}, \citenamefont {Gehl}, \citenamefont {Karl}, \citenamefont {Van Der~Wall}, \citenamefont {Law}, \citenamefont {Blain}, \citenamefont {Stick},\ and\ \citenamefont {McGuinness}}]{kwon2024multi}%
  \BibitemOpen
  \bibfield  {author} {\bibinfo {author} {\bibfnamefont {J.}~\bibnamefont {Kwon}}, \bibinfo {author} {\bibfnamefont {W.~J.}\ \bibnamefont {Setzer}}, \bibinfo {author} {\bibfnamefont {M.}~\bibnamefont {Gehl}}, \bibinfo {author} {\bibfnamefont {N.}~\bibnamefont {Karl}}, \bibinfo {author} {\bibfnamefont {J.}~\bibnamefont {Van Der~Wall}}, \bibinfo {author} {\bibfnamefont {R.}~\bibnamefont {Law}}, \bibinfo {author} {\bibfnamefont {M.~G.}\ \bibnamefont {Blain}}, \bibinfo {author} {\bibfnamefont {D.}~\bibnamefont {Stick}},\ and\ \bibinfo {author} {\bibfnamefont {H.~J.}\ \bibnamefont {McGuinness}},\ }\bibfield  {title} {\bibinfo {title} {Multi-site integrated optical addressing of trapped ions},\ }\href {https://doi.org/https://doi.org/10.1038/s41467-024-47882-5} {\bibfield  {journal} {\bibinfo  {journal} {Nat. Commun.}\ }\textbf {\bibinfo {volume} {15}},\ \bibinfo {pages} {3709} (\bibinfo {year} {2024})}\BibitemShut {NoStop}%
\bibitem [{\citenamefont {Wu}\ \emph {et~al.}(2021)\citenamefont {Wu}, \citenamefont {Wang}, \citenamefont {Han}, \citenamefont {Jiang}, \citenamefont {Song}, \citenamefont {Xia}, \citenamefont {Su},\ and\ \citenamefont {Li}}]{wu2021systematic}%
  \BibitemOpen
  \bibfield  {author} {\bibinfo {author} {\bibfnamefont {J.-L.}\ \bibnamefont {Wu}}, \bibinfo {author} {\bibfnamefont {Y.}~\bibnamefont {Wang}}, \bibinfo {author} {\bibfnamefont {J.-X.}\ \bibnamefont {Han}}, \bibinfo {author} {\bibfnamefont {Y.}~\bibnamefont {Jiang}}, \bibinfo {author} {\bibfnamefont {J.}~\bibnamefont {Song}}, \bibinfo {author} {\bibfnamefont {Y.}~\bibnamefont {Xia}}, \bibinfo {author} {\bibfnamefont {S.-L.}\ \bibnamefont {Su}},\ and\ \bibinfo {author} {\bibfnamefont {W.}~\bibnamefont {Li}},\ }\bibfield  {title} {\bibinfo {title} {Systematic-error-tolerant multiqubit holonomic entangling gates},\ }\href {https://doi.org/https://doi.org/10.1103/PhysRevApplied.16.064031} {\bibfield  {journal} {\bibinfo  {journal} {Phys. Rev. Appl.}\ }\textbf {\bibinfo {volume} {16}},\ \bibinfo {pages} {064031} (\bibinfo {year} {2021})}\BibitemShut {NoStop}%
\bibitem [{\citenamefont {Manetsch}\ \emph {et~al.}(2025)\citenamefont {Manetsch}, \citenamefont {Nomura}, \citenamefont {Bataille}, \citenamefont {Lv}, \citenamefont {Leung},\ and\ \citenamefont {Endres}}]{manetsch2025tweezer}%
  \BibitemOpen
  \bibfield  {author} {\bibinfo {author} {\bibfnamefont {H.~J.}\ \bibnamefont {Manetsch}}, \bibinfo {author} {\bibfnamefont {G.}~\bibnamefont {Nomura}}, \bibinfo {author} {\bibfnamefont {E.}~\bibnamefont {Bataille}}, \bibinfo {author} {\bibfnamefont {X.}~\bibnamefont {Lv}}, \bibinfo {author} {\bibfnamefont {K.~H.}\ \bibnamefont {Leung}},\ and\ \bibinfo {author} {\bibfnamefont {M.}~\bibnamefont {Endres}},\ }\bibfield  {title} {\bibinfo {title} {A tweezer array with 6,100 highly coherent atomic qubits},\ }\href {https://doi.org/https://doi.org/10.1038/s41586-025-09641-4} {\bibfield  {journal} {\bibinfo  {journal} {Nature}\ }\textbf {\bibinfo {volume} {647}},\ \bibinfo {pages} {60} (\bibinfo {year} {2025})}\BibitemShut {NoStop}%
\bibitem [{\citenamefont {Wu}\ \emph {et~al.}(2025)\citenamefont {Wu}, \citenamefont {Wu}, \citenamefont {Guo}, \citenamefont {Liu}, \citenamefont {Su}, \citenamefont {Song}, \citenamefont {Ye},\ and\ \citenamefont {Wang}}]{wu2025quantum}%
  \BibitemOpen
  \bibfield  {author} {\bibinfo {author} {\bibfnamefont {J.}~\bibnamefont {Wu}}, \bibinfo {author} {\bibfnamefont {J.-L.}\ \bibnamefont {Wu}}, \bibinfo {author} {\bibfnamefont {F.-Q.}\ \bibnamefont {Guo}}, \bibinfo {author} {\bibfnamefont {B.-B.}\ \bibnamefont {Liu}}, \bibinfo {author} {\bibfnamefont {S.-L.}\ \bibnamefont {Su}}, \bibinfo {author} {\bibfnamefont {X.-K.}\ \bibnamefont {Song}}, \bibinfo {author} {\bibfnamefont {L.}~\bibnamefont {Ye}},\ and\ \bibinfo {author} {\bibfnamefont {D.}~\bibnamefont {Wang}},\ }\bibfield  {title} {\bibinfo {title} {Quantum computation via {Floquet} tailored {Rydberg} interactions},\ }\href {https://doi.org/https://doi.org/10.1038/s41534-025-01068-z} {\bibfield  {journal} {\bibinfo  {journal} {npj Quantum Inf.}\ }\textbf {\bibinfo {volume} {11}},\ \bibinfo {pages} {118} (\bibinfo {year} {2025})}\BibitemShut {NoStop}%
\bibitem [{\citenamefont {Levine}\ \emph {et~al.}(2019)\citenamefont {Levine}, \citenamefont {Keesling}, \citenamefont {Semeghini}, \citenamefont {Omran}, \citenamefont {Wang}, \citenamefont {Ebadi}, \citenamefont {Bernien}, \citenamefont {Greiner}, \citenamefont {Vuleti{\'c}}, \citenamefont {Pichler} \emph {et~al.}}]{levine2019parallel}%
  \BibitemOpen
  \bibfield  {author} {\bibinfo {author} {\bibfnamefont {H.}~\bibnamefont {Levine}}, \bibinfo {author} {\bibfnamefont {A.}~\bibnamefont {Keesling}}, \bibinfo {author} {\bibfnamefont {G.}~\bibnamefont {Semeghini}}, \bibinfo {author} {\bibfnamefont {A.}~\bibnamefont {Omran}}, \bibinfo {author} {\bibfnamefont {T.~T.}\ \bibnamefont {Wang}}, \bibinfo {author} {\bibfnamefont {S.}~\bibnamefont {Ebadi}}, \bibinfo {author} {\bibfnamefont {H.}~\bibnamefont {Bernien}}, \bibinfo {author} {\bibfnamefont {M.}~\bibnamefont {Greiner}}, \bibinfo {author} {\bibfnamefont {V.}~\bibnamefont {Vuleti{\'c}}}, \bibinfo {author} {\bibfnamefont {H.}~\bibnamefont {Pichler}}, \emph {et~al.},\ }\bibfield  {title} {\bibinfo {title} {Parallel implementation of high-fidelity multiqubit gates with neutral atoms},\ }\href {https://doi.org/https://doi.org/10.1103/PhysRevLett.123.170503} {\bibfield  {journal} {\bibinfo  {journal} {Phys. Rev. Lett.}\ }\textbf {\bibinfo {volume} {123}},\ \bibinfo {pages} {170503} (\bibinfo {year}
  {2019})}\BibitemShut {NoStop}%
\bibitem [{\citenamefont {Evered}\ \emph {et~al.}(2023)\citenamefont {Evered}, \citenamefont {Bluvstein}, \citenamefont {Kalinowski}, \citenamefont {Ebadi}, \citenamefont {Manovitz}, \citenamefont {Zhou}, \citenamefont {Li}, \citenamefont {Geim}, \citenamefont {Wang}, \citenamefont {Maskara} \emph {et~al.}}]{evered2023high}%
  \BibitemOpen
  \bibfield  {author} {\bibinfo {author} {\bibfnamefont {S.~J.}\ \bibnamefont {Evered}}, \bibinfo {author} {\bibfnamefont {D.}~\bibnamefont {Bluvstein}}, \bibinfo {author} {\bibfnamefont {M.}~\bibnamefont {Kalinowski}}, \bibinfo {author} {\bibfnamefont {S.}~\bibnamefont {Ebadi}}, \bibinfo {author} {\bibfnamefont {T.}~\bibnamefont {Manovitz}}, \bibinfo {author} {\bibfnamefont {H.}~\bibnamefont {Zhou}}, \bibinfo {author} {\bibfnamefont {S.~H.}\ \bibnamefont {Li}}, \bibinfo {author} {\bibfnamefont {A.~A.}\ \bibnamefont {Geim}}, \bibinfo {author} {\bibfnamefont {T.~T.}\ \bibnamefont {Wang}}, \bibinfo {author} {\bibfnamefont {N.}~\bibnamefont {Maskara}}, \emph {et~al.},\ }\bibfield  {title} {\bibinfo {title} {High-fidelity parallel entangling gates on a neutral-atom quantum computer},\ }\href {https://doi.org/https://doi.org/10.1038/s41586-023-06481-y} {\bibfield  {journal} {\bibinfo  {journal} {Nature}\ }\textbf {\bibinfo {volume} {622}},\ \bibinfo {pages} {268} (\bibinfo {year} {2023})}\BibitemShut {NoStop}%
\bibitem [{\citenamefont {M.~Farouk}\ \emph {et~al.}(2023)\citenamefont {M.~Farouk}, \citenamefont {Beterov}, \citenamefont {Xu}, \citenamefont {Bergamini},\ and\ \citenamefont {Ryabtsev}}]{m2023parallel}%
  \BibitemOpen
  \bibfield  {author} {\bibinfo {author} {\bibfnamefont {A.}~\bibnamefont {M.~Farouk}}, \bibinfo {author} {\bibfnamefont {I.~I.}\ \bibnamefont {Beterov}}, \bibinfo {author} {\bibfnamefont {P.}~\bibnamefont {Xu}}, \bibinfo {author} {\bibfnamefont {S.}~\bibnamefont {Bergamini}},\ and\ \bibinfo {author} {\bibfnamefont {I.~I.}\ \bibnamefont {Ryabtsev}},\ }\bibfield  {title} {\bibinfo {title} {Parallel implementation of {CNOT}$^\text{N}$ and {C}$_2${NOT}$^2$ gates via homonuclear and heteronuclear {F}{\"o}rster interactions of {Rydberg} atoms},\ }\href {https://doi.org/https://doi.org/10.3390/photonics10111280} {\bibfield  {journal} {\bibinfo  {journal} {Photonics}\ }\textbf {\bibinfo {volume} {10}},\ \bibinfo {pages} {1280} (\bibinfo {year} {2023})}\BibitemShut {NoStop}%
\bibitem [{\citenamefont {Graham}\ \emph {et~al.}(2019)\citenamefont {Graham}, \citenamefont {Kwon}, \citenamefont {Grinkemeyer}, \citenamefont {Marra}, \citenamefont {Jiang}, \citenamefont {Lichtman}, \citenamefont {Sun}, \citenamefont {Ebert},\ and\ \citenamefont {Saffman}}]{graham2019rydberg}%
  \BibitemOpen
  \bibfield  {author} {\bibinfo {author} {\bibfnamefont {T.}~\bibnamefont {Graham}}, \bibinfo {author} {\bibfnamefont {M.}~\bibnamefont {Kwon}}, \bibinfo {author} {\bibfnamefont {B.}~\bibnamefont {Grinkemeyer}}, \bibinfo {author} {\bibfnamefont {Z.}~\bibnamefont {Marra}}, \bibinfo {author} {\bibfnamefont {X.}~\bibnamefont {Jiang}}, \bibinfo {author} {\bibfnamefont {M.}~\bibnamefont {Lichtman}}, \bibinfo {author} {\bibfnamefont {Y.}~\bibnamefont {Sun}}, \bibinfo {author} {\bibfnamefont {M.}~\bibnamefont {Ebert}},\ and\ \bibinfo {author} {\bibfnamefont {M.}~\bibnamefont {Saffman}},\ }\bibfield  {title} {\bibinfo {title} {Rydberg-mediated entanglement in a two-dimensional neutral atom qubit array},\ }\href {https://doi.org/https://doi.org/10.1103/PhysRevLett.123.230501} {\bibfield  {journal} {\bibinfo  {journal} {Phys. Rev. Lett.}\ }\textbf {\bibinfo {volume} {123}},\ \bibinfo {pages} {230501} (\bibinfo {year} {2019})}\BibitemShut {NoStop}%
\bibitem [{\citenamefont {Henriet}\ \emph {et~al.}(2020)\citenamefont {Henriet}, \citenamefont {Beguin}, \citenamefont {Signoles}, \citenamefont {Lahaye}, \citenamefont {Browaeys}, \citenamefont {Reymond},\ and\ \citenamefont {Jurczak}}]{henriet2020quantum}%
  \BibitemOpen
  \bibfield  {author} {\bibinfo {author} {\bibfnamefont {L.}~\bibnamefont {Henriet}}, \bibinfo {author} {\bibfnamefont {L.}~\bibnamefont {Beguin}}, \bibinfo {author} {\bibfnamefont {A.}~\bibnamefont {Signoles}}, \bibinfo {author} {\bibfnamefont {T.}~\bibnamefont {Lahaye}}, \bibinfo {author} {\bibfnamefont {A.}~\bibnamefont {Browaeys}}, \bibinfo {author} {\bibfnamefont {G.-O.}\ \bibnamefont {Reymond}},\ and\ \bibinfo {author} {\bibfnamefont {C.}~\bibnamefont {Jurczak}},\ }\bibfield  {title} {\bibinfo {title} {Quantum computing with neutral atoms},\ }\href {https://doi.org/https://doi.org/10.22331/q-2020-09-21-327} {\bibfield  {journal} {\bibinfo  {journal} {Quantum}\ }\textbf {\bibinfo {volume} {4}},\ \bibinfo {pages} {327} (\bibinfo {year} {2020})}\BibitemShut {NoStop}%
\bibitem [{\citenamefont {Bluvstein}\ \emph {et~al.}(2024)\citenamefont {Bluvstein}, \citenamefont {Evered}, \citenamefont {Geim}, \citenamefont {Li}, \citenamefont {Zhou}, \citenamefont {Manovitz}, \citenamefont {Ebadi}, \citenamefont {Cain}, \citenamefont {Kalinowski}, \citenamefont {Hangleiter} \emph {et~al.}}]{bluvstein2024logical}%
  \BibitemOpen
  \bibfield  {author} {\bibinfo {author} {\bibfnamefont {D.}~\bibnamefont {Bluvstein}}, \bibinfo {author} {\bibfnamefont {S.~J.}\ \bibnamefont {Evered}}, \bibinfo {author} {\bibfnamefont {A.~A.}\ \bibnamefont {Geim}}, \bibinfo {author} {\bibfnamefont {S.~H.}\ \bibnamefont {Li}}, \bibinfo {author} {\bibfnamefont {H.}~\bibnamefont {Zhou}}, \bibinfo {author} {\bibfnamefont {T.}~\bibnamefont {Manovitz}}, \bibinfo {author} {\bibfnamefont {S.}~\bibnamefont {Ebadi}}, \bibinfo {author} {\bibfnamefont {M.}~\bibnamefont {Cain}}, \bibinfo {author} {\bibfnamefont {M.}~\bibnamefont {Kalinowski}}, \bibinfo {author} {\bibfnamefont {D.}~\bibnamefont {Hangleiter}}, \emph {et~al.},\ }\bibfield  {title} {\bibinfo {title} {Logical quantum processor based on reconfigurable atom arrays},\ }\href {https://doi.org/https://doi.org/10.1038/s41586-023-06927-3} {\bibfield  {journal} {\bibinfo  {journal} {Nature}\ }\textbf {\bibinfo {volume} {626}},\ \bibinfo {pages} {58} (\bibinfo {year} {2024})}\BibitemShut {NoStop}%
\bibitem [{\citenamefont {DeMille}(2002)}]{demille2002quantum}%
  \BibitemOpen
  \bibfield  {author} {\bibinfo {author} {\bibfnamefont {D.}~\bibnamefont {DeMille}},\ }\bibfield  {title} {\bibinfo {title} {Quantum computation with trapped polar molecules},\ }\href {https://doi.org/https://doi.org/10.1103/PhysRevLett.88.067901} {\bibfield  {journal} {\bibinfo  {journal} {Phys. Rev. Lett.}\ }\textbf {\bibinfo {volume} {88}},\ \bibinfo {pages} {067901} (\bibinfo {year} {2002})}\BibitemShut {NoStop}%
\bibitem [{\citenamefont {Bao}\ \emph {et~al.}(2023)\citenamefont {Bao}, \citenamefont {Yu}, \citenamefont {Anderegg}, \citenamefont {Chae}, \citenamefont {Ketterle}, \citenamefont {Ni},\ and\ \citenamefont {Doyle}}]{bao2023dipolar}%
  \BibitemOpen
  \bibfield  {author} {\bibinfo {author} {\bibfnamefont {Y.}~\bibnamefont {Bao}}, \bibinfo {author} {\bibfnamefont {S.~S.}\ \bibnamefont {Yu}}, \bibinfo {author} {\bibfnamefont {L.}~\bibnamefont {Anderegg}}, \bibinfo {author} {\bibfnamefont {E.}~\bibnamefont {Chae}}, \bibinfo {author} {\bibfnamefont {W.}~\bibnamefont {Ketterle}}, \bibinfo {author} {\bibfnamefont {K.-K.}\ \bibnamefont {Ni}},\ and\ \bibinfo {author} {\bibfnamefont {J.~M.}\ \bibnamefont {Doyle}},\ }\bibfield  {title} {\bibinfo {title} {Dipolar spin-exchange and entanglement between molecules in an optical tweezer array},\ }\href {https://doi.org/https://doi.org/10.1126/science.adf8999} {\bibfield  {journal} {\bibinfo  {journal} {Science}\ }\textbf {\bibinfo {volume} {382}},\ \bibinfo {pages} {1138} (\bibinfo {year} {2023})}\BibitemShut {NoStop}%
\bibitem [{\citenamefont {Langen}\ \emph {et~al.}(2024)\citenamefont {Langen}, \citenamefont {Valtolina}, \citenamefont {Wang},\ and\ \citenamefont {Ye}}]{langen2024quantum}%
  \BibitemOpen
  \bibfield  {author} {\bibinfo {author} {\bibfnamefont {T.}~\bibnamefont {Langen}}, \bibinfo {author} {\bibfnamefont {G.}~\bibnamefont {Valtolina}}, \bibinfo {author} {\bibfnamefont {D.}~\bibnamefont {Wang}},\ and\ \bibinfo {author} {\bibfnamefont {J.}~\bibnamefont {Ye}},\ }\bibfield  {title} {\bibinfo {title} {Quantum state manipulation and cooling of ultracold molecules},\ }\href {https://doi.org/https://doi.org/10.1038/s41567-024-02423-1} {\bibfield  {journal} {\bibinfo  {journal} {Nat. Phys.}\ }\textbf {\bibinfo {volume} {20}},\ \bibinfo {pages} {702} (\bibinfo {year} {2024})}\BibitemShut {NoStop}%
\bibitem [{\citenamefont {Picard}\ \emph {et~al.}(2025)\citenamefont {Picard}, \citenamefont {Park}, \citenamefont {Patenotte}, \citenamefont {Gebretsadkan}, \citenamefont {Wellnitz}, \citenamefont {Rey},\ and\ \citenamefont {Ni}}]{picard2025entanglement}%
  \BibitemOpen
  \bibfield  {author} {\bibinfo {author} {\bibfnamefont {L.~R.}\ \bibnamefont {Picard}}, \bibinfo {author} {\bibfnamefont {A.~J.}\ \bibnamefont {Park}}, \bibinfo {author} {\bibfnamefont {G.~E.}\ \bibnamefont {Patenotte}}, \bibinfo {author} {\bibfnamefont {S.}~\bibnamefont {Gebretsadkan}}, \bibinfo {author} {\bibfnamefont {D.}~\bibnamefont {Wellnitz}}, \bibinfo {author} {\bibfnamefont {A.~M.}\ \bibnamefont {Rey}},\ and\ \bibinfo {author} {\bibfnamefont {K.-K.}\ \bibnamefont {Ni}},\ }\bibfield  {title} {\bibinfo {title} {Entanglement and {iSWAP} gate between molecular qubits},\ }\href {https://doi.org/https://doi.org/10.1038/s41586-024-08177-3} {\bibfield  {journal} {\bibinfo  {journal} {Nature}\ }\textbf {\bibinfo {volume} {637}},\ \bibinfo {pages} {821} (\bibinfo {year} {2025})}\BibitemShut {NoStop}%
\bibitem [{\citenamefont {Ruttley}\ \emph {et~al.}(2025)\citenamefont {Ruttley}, \citenamefont {Hepworth}, \citenamefont {Guttridge},\ and\ \citenamefont {Cornish}}]{ruttley2025long}%
  \BibitemOpen
  \bibfield  {author} {\bibinfo {author} {\bibfnamefont {D.~K.}\ \bibnamefont {Ruttley}}, \bibinfo {author} {\bibfnamefont {T.~R.}\ \bibnamefont {Hepworth}}, \bibinfo {author} {\bibfnamefont {A.}~\bibnamefont {Guttridge}},\ and\ \bibinfo {author} {\bibfnamefont {S.~L.}\ \bibnamefont {Cornish}},\ }\bibfield  {title} {\bibinfo {title} {Long-lived entanglement of molecules in magic-wavelength optical tweezers},\ }\href {https://doi.org/https://doi.org/10.1038/s41586-024-08365-1} {\bibfield  {journal} {\bibinfo  {journal} {Nature}\ }\textbf {\bibinfo {volume} {637}},\ \bibinfo {pages} {827} (\bibinfo {year} {2025})}\BibitemShut {NoStop}%
\bibitem [{\citenamefont {Wang}\ \emph {et~al.}(2012)\citenamefont {Wang}, \citenamefont {Zhang}, \citenamefont {Tyryshkin}, \citenamefont {Lyon}, \citenamefont {Ager}, \citenamefont {Haller},\ and\ \citenamefont {Dobrovitski}}]{wang2012effect}%
  \BibitemOpen
  \bibfield  {author} {\bibinfo {author} {\bibfnamefont {Z.-H.}\ \bibnamefont {Wang}}, \bibinfo {author} {\bibfnamefont {W.}~\bibnamefont {Zhang}}, \bibinfo {author} {\bibfnamefont {A.~M.}\ \bibnamefont {Tyryshkin}}, \bibinfo {author} {\bibfnamefont {S.~A.}\ \bibnamefont {Lyon}}, \bibinfo {author} {\bibfnamefont {J.}~\bibnamefont {Ager}}, \bibinfo {author} {\bibfnamefont {E.}~\bibnamefont {Haller}},\ and\ \bibinfo {author} {\bibfnamefont {V.~V.}\ \bibnamefont {Dobrovitski}},\ }\bibfield  {title} {\bibinfo {title} {Effect of pulse error accumulation on dynamical decoupling of the electron spins of phosphorus donors in silicon},\ }\href {https://doi.org/https://doi.org/10.1103/PhysRevB.85.085206} {\bibfield  {journal} {\bibinfo  {journal} {Phys. Rev. B}\ }\textbf {\bibinfo {volume} {85}},\ \bibinfo {pages} {085206} (\bibinfo {year} {2012})}\BibitemShut {NoStop}%
\bibitem [{\citenamefont {Souza}\ \emph {et~al.}(2011)\citenamefont {Souza}, \citenamefont {Alvarez},\ and\ \citenamefont {Suter}}]{souza2011robust}%
  \BibitemOpen
  \bibfield  {author} {\bibinfo {author} {\bibfnamefont {A.~M.}\ \bibnamefont {Souza}}, \bibinfo {author} {\bibfnamefont {G.~A.}\ \bibnamefont {Alvarez}},\ and\ \bibinfo {author} {\bibfnamefont {D.}~\bibnamefont {Suter}},\ }\bibfield  {title} {\bibinfo {title} {Robust dynamical decoupling for quantum computing and quantum memory},\ }\href {https://doi.org/https://doi.org/10.1103/PhysRevLett.106.240501} {\bibfield  {journal} {\bibinfo  {journal} {Phys. Rev. Lett.}\ }\textbf {\bibinfo {volume} {106}},\ \bibinfo {pages} {240501} (\bibinfo {year} {2011})}\BibitemShut {NoStop}%
\bibitem [{\citenamefont {Ezzell}\ \emph {et~al.}(2023)\citenamefont {Ezzell}, \citenamefont {Pokharel}, \citenamefont {Tewala}, \citenamefont {Quiroz},\ and\ \citenamefont {Lidar}}]{ezzell2023dynamical}%
  \BibitemOpen
  \bibfield  {author} {\bibinfo {author} {\bibfnamefont {N.}~\bibnamefont {Ezzell}}, \bibinfo {author} {\bibfnamefont {B.}~\bibnamefont {Pokharel}}, \bibinfo {author} {\bibfnamefont {L.}~\bibnamefont {Tewala}}, \bibinfo {author} {\bibfnamefont {G.}~\bibnamefont {Quiroz}},\ and\ \bibinfo {author} {\bibfnamefont {D.~A.}\ \bibnamefont {Lidar}},\ }\bibfield  {title} {\bibinfo {title} {Dynamical decoupling for superconducting qubits: A performance survey},\ }\href {https://doi.org/https://doi.org/10.1103/PhysRevApplied.20.064027} {\bibfield  {journal} {\bibinfo  {journal} {Phys. Rev. Appl.}\ }\textbf {\bibinfo {volume} {20}},\ \bibinfo {pages} {064027} (\bibinfo {year} {2023})}\BibitemShut {NoStop}%
\bibitem [{\citenamefont {Cornish}\ \emph {et~al.}(2024)\citenamefont {Cornish}, \citenamefont {Tarbutt},\ and\ \citenamefont {Hazzard}}]{cornish2024quantum}%
  \BibitemOpen
  \bibfield  {author} {\bibinfo {author} {\bibfnamefont {S.~L.}\ \bibnamefont {Cornish}}, \bibinfo {author} {\bibfnamefont {M.~R.}\ \bibnamefont {Tarbutt}},\ and\ \bibinfo {author} {\bibfnamefont {K.~R.}\ \bibnamefont {Hazzard}},\ }\bibfield  {title} {\bibinfo {title} {Quantum computation and quantum simulation with ultracold molecules},\ }\href {https://doi.org/https://doi.org/10.1038/s41567-024-02453-9} {\bibfield  {journal} {\bibinfo  {journal} {Nat. Phys.}\ }\textbf {\bibinfo {volume} {20}},\ \bibinfo {pages} {730} (\bibinfo {year} {2024})}\BibitemShut {NoStop}%
\bibitem [{\citenamefont {Aymar}\ and\ \citenamefont {Dulieu}(2005)}]{aymar2005calculation}%
  \BibitemOpen
  \bibfield  {author} {\bibinfo {author} {\bibfnamefont {M.}~\bibnamefont {Aymar}}\ and\ \bibinfo {author} {\bibfnamefont {O.}~\bibnamefont {Dulieu}},\ }\bibfield  {title} {\bibinfo {title} {Calculation of accurate permanent dipole moments of the lowest $^{1,3}${$\Sigma$}$^+$ states of heteronuclear alkali dimers using extended basis sets},\ }\href {https://doi.org/https://doi.org/10.1063/1.1903944} {\bibfield  {journal} {\bibinfo  {journal} {J. Chem. Phys.}\ }\textbf {\bibinfo {volume} {122}},\ \bibinfo {pages} {204302} (\bibinfo {year} {2005})}\BibitemShut {NoStop}%
\bibitem [{\citenamefont {Dagdigian}\ and\ \citenamefont {Wharton}(1972)}]{dagdigian1972molecular}%
  \BibitemOpen
  \bibfield  {author} {\bibinfo {author} {\bibfnamefont {P.~J.}\ \bibnamefont {Dagdigian}}\ and\ \bibinfo {author} {\bibfnamefont {L.}~\bibnamefont {Wharton}},\ }\bibfield  {title} {\bibinfo {title} {Molecular beam electric deflection and resonance spectroscopy of the heteronuclear alkali dimers: $^{39}${K}$^7${Li}, {Rb}$^{7}${Li}, $^{39}${K}$^{23}${Na}, {Rb}$^{23}${Na}, and $^{133}${Cs}$^{23}${Na}},\ }\href {https://doi.org/https://doi.org/10.1063/1.1678429} {\bibfield  {journal} {\bibinfo  {journal} {J. Chem. Phys.}\ }\textbf {\bibinfo {volume} {57}},\ \bibinfo {pages} {1487} (\bibinfo {year} {1972})}\BibitemShut {NoStop}%
\bibitem [{\citenamefont {Nielsen}\ and\ \citenamefont {Chuang}(2010)}]{nielsen2010quantum}%
  \BibitemOpen
  \bibfield  {author} {\bibinfo {author} {\bibfnamefont {M.~A.}\ \bibnamefont {Nielsen}}\ and\ \bibinfo {author} {\bibfnamefont {I.~L.}\ \bibnamefont {Chuang}},\ }\href@noop {} {\emph {\bibinfo {title} {Quantum Computation and Quantum Information: 10th Anniversary Edition}}}\ (\bibinfo  {publisher} {Cambridge University Press, Cambridge, England},\ \bibinfo {year} {2010})\BibitemShut {NoStop}%
\bibitem [{\citenamefont {Lee}\ \emph {et~al.}(2017)\citenamefont {Lee}, \citenamefont {Song},\ and\ \citenamefont {Ahn}}]{lee2017single}%
  \BibitemOpen
  \bibfield  {author} {\bibinfo {author} {\bibfnamefont {H.-G.}\ \bibnamefont {Lee}}, \bibinfo {author} {\bibfnamefont {Y.}~\bibnamefont {Song}},\ and\ \bibinfo {author} {\bibfnamefont {J.}~\bibnamefont {Ahn}},\ }\bibfield  {title} {\bibinfo {title} {Single-laser-pulse implementation of arbitrary {ZYZ} rotations of an atomic qubit},\ }\href {https://doi.org/https://doi.org/10.1103/PhysRevA.96.012326} {\bibfield  {journal} {\bibinfo  {journal} {Phys. Rev. A}\ }\textbf {\bibinfo {volume} {96}},\ \bibinfo {pages} {012326} (\bibinfo {year} {2017})}\BibitemShut {NoStop}%
\bibitem [{\citenamefont {Hong}\ \emph {et~al.}(2021)\citenamefont {Hong}, \citenamefont {Fan}, \citenamefont {Shu},\ and\ \citenamefont {Henriksen}}]{hong2021generation}%
  \BibitemOpen
  \bibfield  {author} {\bibinfo {author} {\bibfnamefont {Q.-Q.}\ \bibnamefont {Hong}}, \bibinfo {author} {\bibfnamefont {L.-B.}\ \bibnamefont {Fan}}, \bibinfo {author} {\bibfnamefont {C.-C.}\ \bibnamefont {Shu}},\ and\ \bibinfo {author} {\bibfnamefont {N.~E.}\ \bibnamefont {Henriksen}},\ }\bibfield  {title} {\bibinfo {title} {Generation of maximal three-state field-free molecular orientation with terahertz pulses},\ }\href {https://doi.org/https://doi.org/10.1103/PhysRevA.104.013108} {\bibfield  {journal} {\bibinfo  {journal} {Phys. Rev. A}\ }\textbf {\bibinfo {volume} {104}},\ \bibinfo {pages} {013108} (\bibinfo {year} {2021})}\BibitemShut {NoStop}%
\bibitem [{\citenamefont {Fan}\ \emph {et~al.}(2023)\citenamefont {Fan}, \citenamefont {Shu}, \citenamefont {Dong}, \citenamefont {He}, \citenamefont {Henriksen},\ and\ \citenamefont {Nori}}]{fan2023quantum}%
  \BibitemOpen
  \bibfield  {author} {\bibinfo {author} {\bibfnamefont {L.-B.}\ \bibnamefont {Fan}}, \bibinfo {author} {\bibfnamefont {C.-C.}\ \bibnamefont {Shu}}, \bibinfo {author} {\bibfnamefont {D.}~\bibnamefont {Dong}}, \bibinfo {author} {\bibfnamefont {J.}~\bibnamefont {He}}, \bibinfo {author} {\bibfnamefont {N.~E.}\ \bibnamefont {Henriksen}},\ and\ \bibinfo {author} {\bibfnamefont {F.}~\bibnamefont {Nori}},\ }\bibfield  {title} {\bibinfo {title} {Quantum coherent control of a single molecular-polariton rotation},\ }\href {https://doi.org/https://doi.org/10.1103/PhysRevLett.130.043604} {\bibfield  {journal} {\bibinfo  {journal} {Phys. Rev. Lett.}\ }\textbf {\bibinfo {volume} {130}},\ \bibinfo {pages} {043604} (\bibinfo {year} {2023})}\BibitemShut {NoStop}%
\bibitem [{\citenamefont {Hong}\ \emph {et~al.}(2025)\citenamefont {Hong}, \citenamefont {Dong}, \citenamefont {Henriksen}, \citenamefont {Nori}, \citenamefont {He},\ and\ \citenamefont {Shu}}]{hong2025precise}%
  \BibitemOpen
  \bibfield  {author} {\bibinfo {author} {\bibfnamefont {Q.-Q.}\ \bibnamefont {Hong}}, \bibinfo {author} {\bibfnamefont {D.}~\bibnamefont {Dong}}, \bibinfo {author} {\bibfnamefont {N.~E.}\ \bibnamefont {Henriksen}}, \bibinfo {author} {\bibfnamefont {F.}~\bibnamefont {Nori}}, \bibinfo {author} {\bibfnamefont {J.}~\bibnamefont {He}},\ and\ \bibinfo {author} {\bibfnamefont {C.-C.}\ \bibnamefont {Shu}},\ }\bibfield  {title} {\bibinfo {title} {Precise quantum control of molecular rotation toward a desired orientation},\ }\href {https://doi.org/https://doi.org/10.1103/PhysRevResearch.7.L012049} {\bibfield  {journal} {\bibinfo  {journal} {Phys. Rev. Res.}\ }\textbf {\bibinfo {volume} {7}},\ \bibinfo {pages} {L012049} (\bibinfo {year} {2025})}\BibitemShut {NoStop}%
\bibitem [{\citenamefont {Fan}\ \emph {et~al.}(2025)\citenamefont {Fan}, \citenamefont {Li}, \citenamefont {Chen}, \citenamefont {Zhou}, \citenamefont {Liu},\ and\ \citenamefont {Shu}}]{fan2025maximizing}%
  \BibitemOpen
  \bibfield  {author} {\bibinfo {author} {\bibfnamefont {L.-B.}\ \bibnamefont {Fan}}, \bibinfo {author} {\bibfnamefont {H.-J.}\ \bibnamefont {Li}}, \bibinfo {author} {\bibfnamefont {Q.}~\bibnamefont {Chen}}, \bibinfo {author} {\bibfnamefont {H.}~\bibnamefont {Zhou}}, \bibinfo {author} {\bibfnamefont {H.}~\bibnamefont {Liu}},\ and\ \bibinfo {author} {\bibfnamefont {C.-C.}\ \bibnamefont {Shu}},\ }\bibfield  {title} {\bibinfo {title} {Maximizing orientation of a three-state molecule in a cavity with analytically designed pulses},\ }\href {https://doi.org/https://doi.org/10.1103/PhysRevA.111.033119} {\bibfield  {journal} {\bibinfo  {journal} {Phys. Rev. A}\ }\textbf {\bibinfo {volume} {111}},\ \bibinfo {pages} {033119} (\bibinfo {year} {2025})}\BibitemShut {NoStop}%
\bibitem [{\citenamefont {Hong}\ \emph {et~al.}(2026)\citenamefont {Hong}, \citenamefont {Zhang}, \citenamefont {Shu}, \citenamefont {He}, \citenamefont {Dong},\ and\ \citenamefont {Ding}}]{hong2026precise}%
  \BibitemOpen
  \bibfield  {author} {\bibinfo {author} {\bibfnamefont {Q.-Q.}\ \bibnamefont {Hong}}, \bibinfo {author} {\bibfnamefont {Z.-J.}\ \bibnamefont {Zhang}}, \bibinfo {author} {\bibfnamefont {C.-C.}\ \bibnamefont {Shu}}, \bibinfo {author} {\bibfnamefont {J.}~\bibnamefont {He}}, \bibinfo {author} {\bibfnamefont {D.}~\bibnamefont {Dong}},\ and\ \bibinfo {author} {\bibfnamefont {D.}~\bibnamefont {Ding}},\ }\bibfield  {title} {\bibinfo {title} {Precise quantum control of unidirectional field-free molecular orientation},\ }\href {https://doi.org/https://doi.org/10.1103/4hsj-z8qx} {\bibfield  {journal} {\bibinfo  {journal} {Phys. Rev. A}\ }\textbf {\bibinfo {volume} {113}},\ \bibinfo {pages} {013118} (\bibinfo {year} {2026})}\BibitemShut {NoStop}%
\bibitem [{\citenamefont {Yang}\ \emph {et~al.}(2026)\citenamefont {Yang}, \citenamefont {Hong}, \citenamefont {Ma}, \citenamefont {Ma},\ and\ \citenamefont {Shu}}]{yang2026multilevel}%
  \BibitemOpen
  \bibfield  {author} {\bibinfo {author} {\bibfnamefont {J.}~\bibnamefont {Yang}}, \bibinfo {author} {\bibfnamefont {Q.-Q.}\ \bibnamefont {Hong}}, \bibinfo {author} {\bibfnamefont {S.}~\bibnamefont {Ma}}, \bibinfo {author} {\bibfnamefont {S.-S.}\ \bibnamefont {Ma}},\ and\ \bibinfo {author} {\bibfnamefont {C.-C.}\ \bibnamefont {Shu}},\ }\bibfield  {title} {\bibinfo {title} {Multilevel pulse-area theorem for precise quantum control of molecular vibration and rotation},\ }\href {https://doi.org/https://doi.org/10.1103/mnxs-hjjn} {\bibfield  {journal} {\bibinfo  {journal} {Phys. Rev. A}\ }\textbf {\bibinfo {volume} {113}},\ \bibinfo {pages} {033106} (\bibinfo {year} {2026})}\BibitemShut {NoStop}%
\bibitem [{\citenamefont {Pedersen}\ \emph {et~al.}(2007)\citenamefont {Pedersen}, \citenamefont {M{\o}ller},\ and\ \citenamefont {M{\o}lmer}}]{pedersen2007fidelity}%
  \BibitemOpen
  \bibfield  {author} {\bibinfo {author} {\bibfnamefont {L.~H.}\ \bibnamefont {Pedersen}}, \bibinfo {author} {\bibfnamefont {N.~M.}\ \bibnamefont {M{\o}ller}},\ and\ \bibinfo {author} {\bibfnamefont {K.}~\bibnamefont {M{\o}lmer}},\ }\bibfield  {title} {\bibinfo {title} {Fidelity of quantum operations},\ }\href {https://doi.org/https://doi.org/10.1016/j.physleta.2007.02.069} {\bibfield  {journal} {\bibinfo  {journal} {Phys. Lett. A}\ }\textbf {\bibinfo {volume} {367}},\ \bibinfo {pages} {47} (\bibinfo {year} {2007})}\BibitemShut {NoStop}%
\bibitem [{\citenamefont {Park}\ \emph {et~al.}(2023)\citenamefont {Park}, \citenamefont {Picard}, \citenamefont {Patenotte}, \citenamefont {Zhang}, \citenamefont {Rosenband},\ and\ \citenamefont {Ni}}]{park2023extended}%
  \BibitemOpen
  \bibfield  {author} {\bibinfo {author} {\bibfnamefont {A.~J.}\ \bibnamefont {Park}}, \bibinfo {author} {\bibfnamefont {L.~R.}\ \bibnamefont {Picard}}, \bibinfo {author} {\bibfnamefont {G.~E.}\ \bibnamefont {Patenotte}}, \bibinfo {author} {\bibfnamefont {J.~T.}\ \bibnamefont {Zhang}}, \bibinfo {author} {\bibfnamefont {T.}~\bibnamefont {Rosenband}},\ and\ \bibinfo {author} {\bibfnamefont {K.-K.}\ \bibnamefont {Ni}},\ }\bibfield  {title} {\bibinfo {title} {Extended rotational coherence of polar molecules in an elliptically polarized trap},\ }\href {https://doi.org/https://doi.org/10.1103/PhysRevLett.131.183401} {\bibfield  {journal} {\bibinfo  {journal} {Phys. Rev. Lett.}\ }\textbf {\bibinfo {volume} {131}},\ \bibinfo {pages} {183401} (\bibinfo {year} {2023})}\BibitemShut {NoStop}%
\bibitem [{\citenamefont {Koch}\ \emph {et~al.}(2019)\citenamefont {Koch}, \citenamefont {Lemeshko},\ and\ \citenamefont {Sugny}}]{koch2019quantum}%
  \BibitemOpen
  \bibfield  {author} {\bibinfo {author} {\bibfnamefont {C.~P.}\ \bibnamefont {Koch}}, \bibinfo {author} {\bibfnamefont {M.}~\bibnamefont {Lemeshko}},\ and\ \bibinfo {author} {\bibfnamefont {D.}~\bibnamefont {Sugny}},\ }\bibfield  {title} {\bibinfo {title} {Quantum control of molecular rotation},\ }\href {https://doi.org/https://doi.org/10.1103/RevModPhys.91.035005} {\bibfield  {journal} {\bibinfo  {journal} {Rev. Mod. Phy.}\ }\textbf {\bibinfo {volume} {91}},\ \bibinfo {pages} {035005} (\bibinfo {year} {2019})}\BibitemShut {NoStop}%
\bibitem [{\citenamefont {Mun}\ \emph {et~al.}(2022)\citenamefont {Mun}, \citenamefont {Minemoto}, \citenamefont {Kim},\ and\ \citenamefont {Sakai}}]{mun2022all}%
  \BibitemOpen
  \bibfield  {author} {\bibinfo {author} {\bibfnamefont {J.~H.}\ \bibnamefont {Mun}}, \bibinfo {author} {\bibfnamefont {S.}~\bibnamefont {Minemoto}}, \bibinfo {author} {\bibfnamefont {D.~E.}\ \bibnamefont {Kim}},\ and\ \bibinfo {author} {\bibfnamefont {H.}~\bibnamefont {Sakai}},\ }\bibfield  {title} {\bibinfo {title} {All-optical control of pendular qubit states with nonresonant two-color laser pulses},\ }\href {https://doi.org/https://doi.org/10.1038/s42005-022-01005-y} {\bibfield  {journal} {\bibinfo  {journal} {Commun. Phys.}\ }\textbf {\bibinfo {volume} {5}},\ \bibinfo {pages} {226} (\bibinfo {year} {2022})}\BibitemShut {NoStop}%
\bibitem [{\citenamefont {Hong}\ \emph {et~al.}(2023)\citenamefont {Hong}, \citenamefont {Lian}, \citenamefont {Shu},\ and\ \citenamefont {Henriksen}}]{hong2023quantum}%
  \BibitemOpen
  \bibfield  {author} {\bibinfo {author} {\bibfnamefont {Q.-Q.}\ \bibnamefont {Hong}}, \bibinfo {author} {\bibfnamefont {Z.-Z.}\ \bibnamefont {Lian}}, \bibinfo {author} {\bibfnamefont {C.-C.}\ \bibnamefont {Shu}},\ and\ \bibinfo {author} {\bibfnamefont {N.~E.}\ \bibnamefont {Henriksen}},\ }\bibfield  {title} {\bibinfo {title} {Quantum control of field-free molecular orientation},\ }\href {https://doi.org/https://doi.org/10.1039/D3CP03115B} {\bibfield  {journal} {\bibinfo  {journal} {Phys. Chem. Chem. Phys.}\ }\textbf {\bibinfo {volume} {25}},\ \bibinfo {pages} {32763} (\bibinfo {year} {2023})}\BibitemShut {NoStop}%
\bibitem [{\citenamefont {Tutunnikov}\ \emph {et~al.}(2020)\citenamefont {Tutunnikov}, \citenamefont {Flo{\ss}}, \citenamefont {Gershnabel}, \citenamefont {Brumer}, \citenamefont {Averbukh}, \citenamefont {Milner},\ and\ \citenamefont {Milner}}]{tutunnikov2020observation}%
  \BibitemOpen
  \bibfield  {author} {\bibinfo {author} {\bibfnamefont {I.}~\bibnamefont {Tutunnikov}}, \bibinfo {author} {\bibfnamefont {J.}~\bibnamefont {Flo{\ss}}}, \bibinfo {author} {\bibfnamefont {E.}~\bibnamefont {Gershnabel}}, \bibinfo {author} {\bibfnamefont {P.}~\bibnamefont {Brumer}}, \bibinfo {author} {\bibfnamefont {I.~S.}\ \bibnamefont {Averbukh}}, \bibinfo {author} {\bibfnamefont {A.~A.}\ \bibnamefont {Milner}},\ and\ \bibinfo {author} {\bibfnamefont {V.}~\bibnamefont {Milner}},\ }\bibfield  {title} {\bibinfo {title} {Observation of persistent orientation of chiral molecules by a laser field with twisted polarization},\ }\href {https://doi.org/https://doi.org/10.1103/PhysRevA.101.021403} {\bibfield  {journal} {\bibinfo  {journal} {Phys. Rev. A}\ }\textbf {\bibinfo {volume} {101}},\ \bibinfo {pages} {021403} (\bibinfo {year} {2020})}\BibitemShut {NoStop}%
\bibitem [{\citenamefont {Yang}\ \emph {et~al.}(2016)\citenamefont {Yang}, \citenamefont {Guehr}, \citenamefont {Vecchione}, \citenamefont {Robinson}, \citenamefont {Li}, \citenamefont {Hartmann}, \citenamefont {Shen}, \citenamefont {Coffee}, \citenamefont {Corbett}, \citenamefont {Fry} \emph {et~al.}}]{yang2016diffractive}%
  \BibitemOpen
  \bibfield  {author} {\bibinfo {author} {\bibfnamefont {J.}~\bibnamefont {Yang}}, \bibinfo {author} {\bibfnamefont {M.}~\bibnamefont {Guehr}}, \bibinfo {author} {\bibfnamefont {T.}~\bibnamefont {Vecchione}}, \bibinfo {author} {\bibfnamefont {M.~S.}\ \bibnamefont {Robinson}}, \bibinfo {author} {\bibfnamefont {R.}~\bibnamefont {Li}}, \bibinfo {author} {\bibfnamefont {N.}~\bibnamefont {Hartmann}}, \bibinfo {author} {\bibfnamefont {X.}~\bibnamefont {Shen}}, \bibinfo {author} {\bibfnamefont {R.}~\bibnamefont {Coffee}}, \bibinfo {author} {\bibfnamefont {J.}~\bibnamefont {Corbett}}, \bibinfo {author} {\bibfnamefont {A.}~\bibnamefont {Fry}}, \emph {et~al.},\ }\bibfield  {title} {\bibinfo {title} {Diffractive imaging of a rotational wavepacket in nitrogen molecules with femtosecond megaelectronvolt electron pulses},\ }\href {https://doi.org/https://doi.org/10.1038/ncomms11232} {\bibfield  {journal} {\bibinfo  {journal} {Nat. Commun.}\ }\textbf {\bibinfo {volume} {7}},\ \bibinfo {pages} {11232} (\bibinfo {year}
  {2016})}\BibitemShut {NoStop}%
\bibitem [{\citenamefont {Trippel}\ \emph {et~al.}(2015)\citenamefont {Trippel}, \citenamefont {Mullins}, \citenamefont {M{\"u}ller}, \citenamefont {Kienitz}, \citenamefont {Gonz{\'a}lez-F{\'e}rez},\ and\ \citenamefont {K{\"u}pper}}]{trippel2015two}%
  \BibitemOpen
  \bibfield  {author} {\bibinfo {author} {\bibfnamefont {S.}~\bibnamefont {Trippel}}, \bibinfo {author} {\bibfnamefont {T.}~\bibnamefont {Mullins}}, \bibinfo {author} {\bibfnamefont {N.~L.}\ \bibnamefont {M{\"u}ller}}, \bibinfo {author} {\bibfnamefont {J.~S.}\ \bibnamefont {Kienitz}}, \bibinfo {author} {\bibfnamefont {R.}~\bibnamefont {Gonz{\'a}lez-F{\'e}rez}},\ and\ \bibinfo {author} {\bibfnamefont {J.}~\bibnamefont {K{\"u}pper}},\ }\bibfield  {title} {\bibinfo {title} {Two-state wave packet for strong field-free molecular orientation},\ }\href {https://doi.org/https://doi.org/10.1103/PhysRevLett.114.103003} {\bibfield  {journal} {\bibinfo  {journal} {Phys. Rev. Lett.}\ }\textbf {\bibinfo {volume} {114}},\ \bibinfo {pages} {103003} (\bibinfo {year} {2015})}\BibitemShut {NoStop}%
\bibitem [{\citenamefont {Renard}\ \emph {et~al.}(2003)\citenamefont {Renard}, \citenamefont {Renard}, \citenamefont {Gu{\'e}rin}, \citenamefont {Pashayan}, \citenamefont {Lavorel}, \citenamefont {Faucher},\ and\ \citenamefont {Jauslin}}]{renard2003postpulse}%
  \BibitemOpen
  \bibfield  {author} {\bibinfo {author} {\bibfnamefont {V.}~\bibnamefont {Renard}}, \bibinfo {author} {\bibfnamefont {M.}~\bibnamefont {Renard}}, \bibinfo {author} {\bibfnamefont {S.}~\bibnamefont {Gu{\'e}rin}}, \bibinfo {author} {\bibfnamefont {Y.}~\bibnamefont {Pashayan}}, \bibinfo {author} {\bibfnamefont {B.}~\bibnamefont {Lavorel}}, \bibinfo {author} {\bibfnamefont {O.}~\bibnamefont {Faucher}},\ and\ \bibinfo {author} {\bibfnamefont {H.-R.}\ \bibnamefont {Jauslin}},\ }\bibfield  {title} {\bibinfo {title} {Postpulse molecular alignment measured by a weak field polarization technique},\ }\href {https://doi.org/https://doi.org/10.1103/PhysRevLett.90.153601} {\bibfield  {journal} {\bibinfo  {journal} {Phys. Rev. Lett.}\ }\textbf {\bibinfo {volume} {90}},\ \bibinfo {pages} {153601} (\bibinfo {year} {2003})}\BibitemShut {NoStop}%
\bibitem [{\citenamefont {Peng}\ \emph {et~al.}(2015)\citenamefont {Peng}, \citenamefont {Bai}, \citenamefont {Li},\ and\ \citenamefont {Liu}}]{peng2015measurement}%
  \BibitemOpen
  \bibfield  {author} {\bibinfo {author} {\bibfnamefont {P.}~\bibnamefont {Peng}}, \bibinfo {author} {\bibfnamefont {Y.}~\bibnamefont {Bai}}, \bibinfo {author} {\bibfnamefont {N.}~\bibnamefont {Li}},\ and\ \bibinfo {author} {\bibfnamefont {P.}~\bibnamefont {Liu}},\ }\bibfield  {title} {\bibinfo {title} {Measurement of field-free molecular alignment by balanced weak field polarization technique},\ }\href {https://doi.org/10.1063/1.4937476} {\bibfield  {journal} {\bibinfo  {journal} {AIP Adv.}\ }\textbf {\bibinfo {volume} {5}},\ \bibinfo {pages} {127205} (\bibinfo {year} {2015})}\BibitemShut {NoStop}%
\bibitem [{\citenamefont {Lian}\ \emph {et~al.}(2023)\citenamefont {Lian}, \citenamefont {Chen}, \citenamefont {Li}, \citenamefont {Shu},\ and\ \citenamefont {Hu}}]{lian2023visualizing}%
  \BibitemOpen
  \bibfield  {author} {\bibinfo {author} {\bibfnamefont {Z.}~\bibnamefont {Lian}}, \bibinfo {author} {\bibfnamefont {Z.}~\bibnamefont {Chen}}, \bibinfo {author} {\bibfnamefont {J.}~\bibnamefont {Li}}, \bibinfo {author} {\bibfnamefont {C.-C.}\ \bibnamefont {Shu}},\ and\ \bibinfo {author} {\bibfnamefont {Z.}~\bibnamefont {Hu}},\ }\bibfield  {title} {\bibinfo {title} {Visualizing molecular unidirectional rotation by a rotated weak-field polarization technique},\ }\href {https://doi.org/https://doi.org/10.1103/PhysRevA.108.063108} {\bibfield  {journal} {\bibinfo  {journal} {Phys. Rev. A}\ }\textbf {\bibinfo {volume} {108}},\ \bibinfo {pages} {063108} (\bibinfo {year} {2023})}\BibitemShut {NoStop}%
\bibitem [{\citenamefont {Cairncross}\ \emph {et~al.}(2021)\citenamefont {Cairncross}, \citenamefont {Zhang}, \citenamefont {Picard}, \citenamefont {Yu}, \citenamefont {Wang},\ and\ \citenamefont {Ni}}]{cairncross2021assembly}%
  \BibitemOpen
  \bibfield  {author} {\bibinfo {author} {\bibfnamefont {W.~B.}\ \bibnamefont {Cairncross}}, \bibinfo {author} {\bibfnamefont {J.~T.}\ \bibnamefont {Zhang}}, \bibinfo {author} {\bibfnamefont {L.~R.}\ \bibnamefont {Picard}}, \bibinfo {author} {\bibfnamefont {Y.}~\bibnamefont {Yu}}, \bibinfo {author} {\bibfnamefont {K.}~\bibnamefont {Wang}},\ and\ \bibinfo {author} {\bibfnamefont {K.-K.}\ \bibnamefont {Ni}},\ }\bibfield  {title} {\bibinfo {title} {Assembly of a rovibrational ground state molecule in an optical tweezer},\ }\href {https://doi.org/https://doi.org/10.1103/PhysRevLett.126.123402} {\bibfield  {journal} {\bibinfo  {journal} {Phys. Rev. Lett.}\ }\textbf {\bibinfo {volume} {126}},\ \bibinfo {pages} {123402} (\bibinfo {year} {2021})}\BibitemShut {NoStop}%
\bibitem [{\citenamefont {Yoshihara}\ \emph {et~al.}(2014)\citenamefont {Yoshihara}, \citenamefont {Nakamura}, \citenamefont {Yan}, \citenamefont {Gustavsson}, \citenamefont {Bylander}, \citenamefont {Oliver},\ and\ \citenamefont {Tsai}}]{yoshihara2014flux}%
  \BibitemOpen
  \bibfield  {author} {\bibinfo {author} {\bibfnamefont {F.}~\bibnamefont {Yoshihara}}, \bibinfo {author} {\bibfnamefont {Y.}~\bibnamefont {Nakamura}}, \bibinfo {author} {\bibfnamefont {F.}~\bibnamefont {Yan}}, \bibinfo {author} {\bibfnamefont {S.}~\bibnamefont {Gustavsson}}, \bibinfo {author} {\bibfnamefont {J.}~\bibnamefont {Bylander}}, \bibinfo {author} {\bibfnamefont {W.~D.}\ \bibnamefont {Oliver}},\ and\ \bibinfo {author} {\bibfnamefont {J.-S.}\ \bibnamefont {Tsai}},\ }\bibfield  {title} {\bibinfo {title} {Flux qubit noise spectroscopy using {Rabi} oscillations under strong driving conditions},\ }\href {https://doi.org/https://doi.org/10.1103/PhysRevB.89.020503} {\bibfield  {journal} {\bibinfo  {journal} {Phys. Rev. B}\ }\textbf {\bibinfo {volume} {89}},\ \bibinfo {pages} {020503(R)} (\bibinfo {year} {2014})}\BibitemShut {NoStop}%
\bibitem [{\citenamefont {Bardin}\ \emph {et~al.}(2021)\citenamefont {Bardin}, \citenamefont {Slichter},\ and\ \citenamefont {Reilly}}]{bardin2021microwaves}%
  \BibitemOpen
  \bibfield  {author} {\bibinfo {author} {\bibfnamefont {J.~C.}\ \bibnamefont {Bardin}}, \bibinfo {author} {\bibfnamefont {D.~H.}\ \bibnamefont {Slichter}},\ and\ \bibinfo {author} {\bibfnamefont {D.~J.}\ \bibnamefont {Reilly}},\ }\bibfield  {title} {\bibinfo {title} {Microwaves in quantum computing},\ }\href {https://doi.org/https://doi.org/10.1109/JMW.2020.3034071} {\bibfield  {journal} {\bibinfo  {journal} {IEEE J. Microwaves}\ }\textbf {\bibinfo {volume} {1}},\ \bibinfo {pages} {403} (\bibinfo {year} {2021})}\BibitemShut {NoStop}%
\bibitem [{\citenamefont {Cronin}\ \emph {et~al.}(2009)\citenamefont {Cronin}, \citenamefont {Schmiedmayer},\ and\ \citenamefont {Pritchard}}]{cronin2009optics}%
  \BibitemOpen
  \bibfield  {author} {\bibinfo {author} {\bibfnamefont {A.~D.}\ \bibnamefont {Cronin}}, \bibinfo {author} {\bibfnamefont {J.}~\bibnamefont {Schmiedmayer}},\ and\ \bibinfo {author} {\bibfnamefont {D.~E.}\ \bibnamefont {Pritchard}},\ }\bibfield  {title} {\bibinfo {title} {Optics and interferometry with atoms and molecules},\ }\href {https://doi.org/https://doi.org/10.1103/RevModPhys.81.1051} {\bibfield  {journal} {\bibinfo  {journal} {Rev. Mod. Phys.}\ }\textbf {\bibinfo {volume} {81}},\ \bibinfo {pages} {1051} (\bibinfo {year} {2009})}\BibitemShut {NoStop}%
\bibitem [{\citenamefont {Borkowski}(2018)}]{borkowski2018optical}%
  \BibitemOpen
  \bibfield  {author} {\bibinfo {author} {\bibfnamefont {M.}~\bibnamefont {Borkowski}},\ }\bibfield  {title} {\bibinfo {title} {Optical lattice clocks with weakly bound molecules},\ }\href {https://doi.org/https://doi.org/10.1103/PhysRevLett.120.083202} {\bibfield  {journal} {\bibinfo  {journal} {Phys. Rev. Lett.}\ }\textbf {\bibinfo {volume} {120}},\ \bibinfo {pages} {083202} (\bibinfo {year} {2018})}\BibitemShut {NoStop}%
\bibitem [{\citenamefont {Zhang}\ \emph {et~al.}(2022)\citenamefont {Zhang}, \citenamefont {Picard}, \citenamefont {Cairncross}, \citenamefont {Wang}, \citenamefont {Yu}, \citenamefont {Fang},\ and\ \citenamefont {Ni}}]{zhang2022optical}%
  \BibitemOpen
  \bibfield  {author} {\bibinfo {author} {\bibfnamefont {J.~T.}\ \bibnamefont {Zhang}}, \bibinfo {author} {\bibfnamefont {L.~R.}\ \bibnamefont {Picard}}, \bibinfo {author} {\bibfnamefont {W.~B.}\ \bibnamefont {Cairncross}}, \bibinfo {author} {\bibfnamefont {K.}~\bibnamefont {Wang}}, \bibinfo {author} {\bibfnamefont {Y.}~\bibnamefont {Yu}}, \bibinfo {author} {\bibfnamefont {F.}~\bibnamefont {Fang}},\ and\ \bibinfo {author} {\bibfnamefont {K.-K.}\ \bibnamefont {Ni}},\ }\bibfield  {title} {\bibinfo {title} {An optical tweezer array of ground-state polar molecules},\ }\href {https://doi.org/10.1088/2058-9565/ac676c} {\bibfield  {journal} {\bibinfo  {journal} {Quantum Sci. Technol.}\ }\textbf {\bibinfo {volume} {7}},\ \bibinfo {pages} {035006} (\bibinfo {year} {2022})}\BibitemShut {NoStop}%
\bibitem [{\citenamefont {Neyenhuis}\ \emph {et~al.}(2012)\citenamefont {Neyenhuis}, \citenamefont {Yan}, \citenamefont {Moses}, \citenamefont {Covey}, \citenamefont {Chotia}, \citenamefont {Petrov}, \citenamefont {Kotochigova}, \citenamefont {Ye},\ and\ \citenamefont {Jin}}]{neyenhuis2012anisotropic}%
  \BibitemOpen
  \bibfield  {author} {\bibinfo {author} {\bibfnamefont {B.}~\bibnamefont {Neyenhuis}}, \bibinfo {author} {\bibfnamefont {B.}~\bibnamefont {Yan}}, \bibinfo {author} {\bibfnamefont {S.}~\bibnamefont {Moses}}, \bibinfo {author} {\bibfnamefont {J.}~\bibnamefont {Covey}}, \bibinfo {author} {\bibfnamefont {A.}~\bibnamefont {Chotia}}, \bibinfo {author} {\bibfnamefont {A.}~\bibnamefont {Petrov}}, \bibinfo {author} {\bibfnamefont {S.}~\bibnamefont {Kotochigova}}, \bibinfo {author} {\bibfnamefont {J.}~\bibnamefont {Ye}},\ and\ \bibinfo {author} {\bibfnamefont {D.}~\bibnamefont {Jin}},\ }\bibfield  {title} {\bibinfo {title} {Anisotropic polarizability of ultracold polar $^{40}${K}$^{87}${Rb} molecules},\ }\href {https://doi.org/https://doi.org/10.1103/PhysRevLett.109.230403} {\bibfield  {journal} {\bibinfo  {journal} {Phys. Rev. Lett.}\ }\textbf {\bibinfo {volume} {109}},\ \bibinfo {pages} {230403} (\bibinfo {year} {2012})}\BibitemShut {NoStop}%
\bibitem [{\citenamefont {Zhang}\ \emph {et~al.}(2024)\citenamefont {Zhang}, \citenamefont {Yuan}, \citenamefont {Bigagli}, \citenamefont {Warner}, \citenamefont {Stevenson},\ and\ \citenamefont {Will}}]{zhang2024dressed}%
  \BibitemOpen
  \bibfield  {author} {\bibinfo {author} {\bibfnamefont {S.}~\bibnamefont {Zhang}}, \bibinfo {author} {\bibfnamefont {W.}~\bibnamefont {Yuan}}, \bibinfo {author} {\bibfnamefont {N.}~\bibnamefont {Bigagli}}, \bibinfo {author} {\bibfnamefont {C.}~\bibnamefont {Warner}}, \bibinfo {author} {\bibfnamefont {I.}~\bibnamefont {Stevenson}},\ and\ \bibinfo {author} {\bibfnamefont {S.}~\bibnamefont {Will}},\ }\bibfield  {title} {\bibinfo {title} {Dressed-state spectroscopy and magic trapping of microwave-shielded {NaCs} molecules},\ }\href {https://doi.org/https://doi.org/10.1103/PhysRevLett.133.263401} {\bibfield  {journal} {\bibinfo  {journal} {Phys. Rev. Lett.}\ }\textbf {\bibinfo {volume} {133}},\ \bibinfo {pages} {263401} (\bibinfo {year} {2024})}\BibitemShut {NoStop}%
\bibitem [{\citenamefont {See{\ss}elberg}\ \emph {et~al.}(2018)\citenamefont {See{\ss}elberg}, \citenamefont {Luo}, \citenamefont {Li}, \citenamefont {Bause}, \citenamefont {Kotochigova}, \citenamefont {Bloch},\ and\ \citenamefont {Gohle}}]{seesselberg2018extending}%
  \BibitemOpen
  \bibfield  {author} {\bibinfo {author} {\bibfnamefont {F.}~\bibnamefont {See{\ss}elberg}}, \bibinfo {author} {\bibfnamefont {X.-Y.}\ \bibnamefont {Luo}}, \bibinfo {author} {\bibfnamefont {M.}~\bibnamefont {Li}}, \bibinfo {author} {\bibfnamefont {R.}~\bibnamefont {Bause}}, \bibinfo {author} {\bibfnamefont {S.}~\bibnamefont {Kotochigova}}, \bibinfo {author} {\bibfnamefont {I.}~\bibnamefont {Bloch}},\ and\ \bibinfo {author} {\bibfnamefont {C.}~\bibnamefont {Gohle}},\ }\bibfield  {title} {\bibinfo {title} {Extending rotational coherence of interacting polar molecules in a spin-decoupled magic trap},\ }\href {https://doi.org/https://doi.org/10.1103/PhysRevLett.121.253401} {\bibfield  {journal} {\bibinfo  {journal} {Phys. Rev. Lett.}\ }\textbf {\bibinfo {volume} {121}},\ \bibinfo {pages} {253401} (\bibinfo {year} {2018})}\BibitemShut {NoStop}%
\bibitem [{\citenamefont {Burchesky}\ \emph {et~al.}(2021)\citenamefont {Burchesky}, \citenamefont {Anderegg}, \citenamefont {Bao}, \citenamefont {Yu}, \citenamefont {Chae}, \citenamefont {Ketterle}, \citenamefont {Ni},\ and\ \citenamefont {Doyle}}]{burchesky2021rotational}%
  \BibitemOpen
  \bibfield  {author} {\bibinfo {author} {\bibfnamefont {S.}~\bibnamefont {Burchesky}}, \bibinfo {author} {\bibfnamefont {L.}~\bibnamefont {Anderegg}}, \bibinfo {author} {\bibfnamefont {Y.}~\bibnamefont {Bao}}, \bibinfo {author} {\bibfnamefont {S.~S.}\ \bibnamefont {Yu}}, \bibinfo {author} {\bibfnamefont {E.}~\bibnamefont {Chae}}, \bibinfo {author} {\bibfnamefont {W.}~\bibnamefont {Ketterle}}, \bibinfo {author} {\bibfnamefont {K.-K.}\ \bibnamefont {Ni}},\ and\ \bibinfo {author} {\bibfnamefont {J.~M.}\ \bibnamefont {Doyle}},\ }\bibfield  {title} {\bibinfo {title} {Rotational coherence times of polar molecules in optical tweezers},\ }\href {https://doi.org/https://doi.org/10.1103/PhysRevLett.127.123202} {\bibfield  {journal} {\bibinfo  {journal} {Phys. Rev. Lett.}\ }\textbf {\bibinfo {volume} {127}},\ \bibinfo {pages} {123202} (\bibinfo {year} {2021})}\BibitemShut {NoStop}%
\end{thebibliography}

%

\end{document}